\documentclass[%
 reprint,
 fleqn,
 floatfix,
 amsmath,amssymb,
 aps,
]{revtex4-2}

\usepackage{graphicx}
\usepackage{dcolumn}
\usepackage{bm}
\usepackage{multirow}
\usepackage{booktabs}
\usepackage{makecell}
\usepackage{adjustbox}
\usepackage{rotating}
\usepackage{placeins}
\usepackage{morefloats}
\usepackage{xcolor}
\PassOptionsToPackage{hyphens}{url}
\usepackage{hyperref}

\extrafloats{80}

\hypersetup{colorlinks=true,linkcolor=blue,citecolor=blue,urlcolor=blue}

\makeatletter
\let\auto@bib@innerbib\relax
\makeatother

\begin{document}

\raggedbottom

\title{Earthquake magnitudes depend on seismic history, as revealed by a neural network analysis}

\author{Neri Berman}
\email{neriberman@gmail.com}
\affiliation{Department of Physics, Tel-Aviv University, Tel-Aviv, Israel}
\affiliation{Google Research, Google, Tel-Aviv, Israel}

\author{Oleg Zlydenko}
\affiliation{Google Research, Google, Tel-Aviv, Israel}

\author{Oren Gilon}
\affiliation{Google Research, Google, Tel-Aviv, Israel}

\author{Yossi Matias}
\affiliation{Google Research, Google, Tel-Aviv, Israel}

\author{Yohai Bar-Sinai}
\email{ybarsinai@gmail.com}
\affiliation{Department of Physics, Tel-Aviv University, Tel-Aviv, Israel}
\affiliation{Racah Institute of Physics, The Hebrew University of Jerusalem, Jerusalem, Israel}
\affiliation{The Rachel and Selim Benin School of Computer Science and Engineering, The Hebrew University of Jerusalem, Jerusalem, Israel}

\date{\today}

\begin{abstract}
    Earthquake occurrence is notoriously difficult to predict.
    While some aspects of their spatiotemporal statistics can be relatively well captured by point-process models, very little is known regarding the magnitude of future events, and it is deeply debated whether it is possible to predict the magnitude of an earthquake before it starts.
    Most operational forecasting models assume that earthquake magnitudes follow a time-independent Gutenberg-Richter (GR) distribution, effectively treating magnitudes as independent of seismic history.
    We address this fundamental question by demonstrating that standard hypocenter catalogs carry information about future earthquake magnitudes, making them more predictable than previously considered.
    We present MAGNET (MAGnitude Neural EsTimation model), which uses a multi-encoder neural network architecture with LSTM units to process spatiotemporal patterns in seismic history.
    By analyzing hypocenter locations, occurrence times, and magnitudes of past events, MAGNET generates probabilistic magnitude forecasts that demonstrate information gains in predicting magnitudes of future events over GR-based models, after controlling for detection artifacts.
    Our model achieves an information gain of approximately 0.07 bit per earthquake on average over the GR benchmark in Southern California, Japan, and New Zealand catalogs, with this advantage persisting.
    These results demonstrate that hypocentral earthquake catalogs contain extractable information about future magnitudes, challenging the conventional separability assumption in earthquake forecasting and offering new approaches for seismic hazard assessment.
\end{abstract}

\maketitle
\section{Introduction} \label{sec:introduction}
Earthquakes are notoriously unpredictable, and forecasting seismicity is a long-standing scientific and technological challenge, often deemed unrealistic due to the inherent complexity of earthquake processes and the scarcity of near-field data~\cite{bernard_earthquake_1999, geller_earthquakes_1997}.
Research since the late 19th century has provided much phenomenological insight about the spatiotemporal statistics of earthquakes, including various marginal distributions~\cite{gutenberg_frequency_1944, kagan_seismic_2002}, scaling relations \cite{bak_earthquakes_1989, dascher-cousineau_what_2020, kagan_aftershock_2002, utsu_centenary_1995} and characteristics of both spatial and temporal clustering \cite{omori_after-shocks_1894, kagan_short-term_2004, ben-zion_localization_2020, devries_deep_2018, king_static_1994}.
Clearly, these insights can be used to quantitatively inform us about future seismicity based on recent history.
For example, Omori's law tells us that after big earthquakes we should expect an increase in the local seismicity rate~\cite{omori_after-shocks_1894}.
Such laws have been incorporated into a variety of forecasting models, which are operationally used today\cite{ogata_statistical_1988, hardebeck_aftershock_2024, jordan_operational_2011, stirling_national_2012}.

However, these statistical relations primarily describe the rate and locations of earthquakes, with limited insight as to the dependence of earthquake magnitudes on seismic history.
The fundamental question of magnitude predictability remains unresolved, and
it is still debated whether it is possible to determine the magnitude of an earthquake before it starts~\cite{kagan_seismic_2002, ogata_exploring_2018}, or even during the initial stages of rupture~\cite{ellsworth_seismic_1995, meier_evidence_2016}.
Multiple previous studies have failed to find robust correlations between earthquake magnitude and seismic history by performing stochastic declustering, excluding data with short-time incompleteness and performing random permutation tests on the magnitudes \cite{petrillo_verifying_2023, taroni_are_2024, davidsen_are_2011}; these works concluded that any observed dependencies were likely spurious artifacts of catalog incompleteness.
Indeed, due to the spatial complexity of the elastic fields, faulting patterns and lithology, the nonlinearities of the rupture process and the complicated interaction between them all, it is not far-fetched to assume that determining the magnitude of an event requires a detailed, perhaps microscopic, knowledge of the system's state. \cite{mignan_seismicity_2012, mignan_debate_2014, picozzi_preparatory_2023, martinez-alvarez_determining_2013,raub_variations_2017, latour_characterization_2013-1, rouet-leduc_machine_2017-2, venegas-aravena_large_2025, zaccagnino_are_2024}
Yet, the inherent complexity of the physical system and scarcity of relevant data often make such physical knowledge inaccessible, leading many to treat seismicity as effectively stochastic. \cite{martinez-garzon_cascade_2024, picozzi_preparatory_2023, ogata_comparing_2014}

Consequently, with the exception of several long-term, event-recurrence-based forecasting models~\cite{wang_earthquake_2024,gerstenberger_seismicity_2024,field_uniform_2014, rhoades_long-range_2004}, state-of-the-art operational forecasting models adopt the simplifying assumption that earthquake magnitudes follow a constant, or slowly changing, Gutenberg-Richter distribution (GR) which characterizes the marginal magnitude statistics over large regions and extended time periods.
This is referred to as the ``separability assumption'':\cite{ogata_statistics_2017, schoenberg_testing_2004} the distribution of earthquake magnitudes is statistically independent of their locations and times.
Mathematically, this means that a conditional intensity function $\lambda$ describing the probability of locations and magnitudes of earthquakes can be separated as $\lambda(t,x,y,M)=\lambda_1(t,x,y)\lambda_2(M)$.
This separability assumption is a core component of common operational forecasting systems, such as the Epidemic-Type Aftershock Sequence (ETAS) model. Consequently, the magnitude-prediction component, $\lambda_2(M)$, of such models is functionally equivalent to the Gutenberg-Richter benchmarks evaluated in this study. \cite{mizrahi_effect_2021-1, ogata_statistical_1988, ogata_statistics_2017}

It should be noted that several deviations from stationarity of the magnitude distribution have been documented: slow changes in the instantaneous magnitude distribution \cite{gulia_effect_2018, gulia_real-time_2019, nandan_magnitude_2019,tormann_systematic_2014,helmstetter_adaptive_2014} are known; large aftershocks tend to occur at greater distances from the mainshock rupture centroid \cite{van_der_elst_larger_2015};
and some faults are known to show recurrent earthquake with a typical magnitude \cite{uchida_repeating_2019, gulia_effect_2018, gulia_real-time_2019, nandan_magnitude_2019,tormann_systematic_2014,helmstetter_adaptive_2014} and recurrence time.
In addition, some studies have shown circumstantial evidence that earthquake magnitudes may not follow a strictly stationary distribution, e.g.~by showing that statistics are not invariant to permutation of the order, or by showing that the maximal magnitude in a given time frame can be predicted with a better-than-random performance~\cite{xiong_seismic_2023, corral_comment_2005, spassiani_exploring_2016, lippiello_positive_2024, lippiello_positive_2024-1, lippiello_influence_2008, shcherbakov_forecasting_2019, panakkat_neural_2007}.

Still, a direct assessment of magnitude predictability is challenging due to data scarcity and various measurement limitations, the most challenging of which is ``short-time incompleteness'' (STI): the fact that immediately following a large event the signal of subsequent small events might be buried in the coda of the mainshock~\cite{kagan_short-term_2004, stockman_forecasting_2023}.
Thus, during this time smaller events are less likely to be included in the catalog not because they did not occur, but rather because they were not detected.
Statistical models for magnitude prediction may show information gain due to this artifact, which is challenging remove from the data.
While STI may be treated as an advantage for magnitude prediction in some contexts, e.g.~Stockman et. al.~\cite{stockman_forecasting_2023}, it serves as a confounding factor when the objective is to model the intrinsic predictability of underlying physical processes.
For example, because smaller events are systematically missed while larger ones remain recorded in the immediate aftermath of a large event, statistical models may misinterpret this relative lack of small magnitudes as a genuine physical trend toward larger magnitudes, effectively mimicking a temporary shift and skewness in magnitude distribution, rather than learning the true underlying distribution of events\cite{lippiello_positive_2024,taroni_are_2024,helmstetter_comparison_2006}.
Still, a robust model demonstrating information gain over a stationary or quasi-stationary GR benchmark in predicting the magnitude of a specific future event is lacking~\cite{ogata_exploring_2018, stockman_forecasting_2023}.

The main goal of this paper is to ask this question directly: is it possible to extract any information about the magnitude of a \emph{specific} future earthquake from regional seismic history? A positive answer would have two important consequences. From a fundamental point of view, it demonstrates that earthquake magnitudes are not inherently unpredictable, which should inform our understanding of the underlying physics of earthquakes.
Second, it suggests that the separability assumption, which is widely applied in operational earthquake forecasting, may be replaced by a more nuanced model that incorporates the seismic history into the magnitude prediction. This may lead to improved forecasting models, and potentially be used to identify precursory signals.

To this end, we construct MAGnitude Neural EsTimation model, \nobreak MAGNET, a neural-based model that predicts the magnitude of a given earthquake given the short and long term seismic history prior to its occurrence, and the timing and location of a specific event.
Our model is trained on seismic catalogs containing hypocenter locations, times, and magnitudes of past earthquakes.
Importantly, explicitly providing the spatiotemporal coordinates of the event is central to our methodology, as it decouples the problem of magnitude predictability from all other spatiotemporal clustering effects.
If our model performs better than a draw from the GR distribution or its variants, as we will indeed demonstrate is the case, we assert that partial information about the magnitude of a \textit{specific future earthquake} is extractable from cataloged properties alone.

The remainder of the manuscript is organized as follows.
In the Methodology section, we present the multi-encoder neural architecture of MAGNET and the probabilistic framework used for the performance of magnitude estimation, including our protocols for benchmarking and strategies for STI mitigation.
In the Results, we demonstrate that MAGNET achieves consistent information gain over all tested benchmarks across three geophysically diverse regions: Southern California, New Zealand, and Japan.
Finally, in the Implications and Outlook section, we discuss the physical interpretation of these findings and their potential integration into early warning and operational forecasting systems.
Additional details regarding our analysis protocols, expanded dataset descriptions, and exhaustive performance comparisons against multiple benchmarks are provided in the Methods and Supplementary Information sections.

\begin{figure*}[ht!]
    \centering
    \includegraphics[width=1\textwidth]{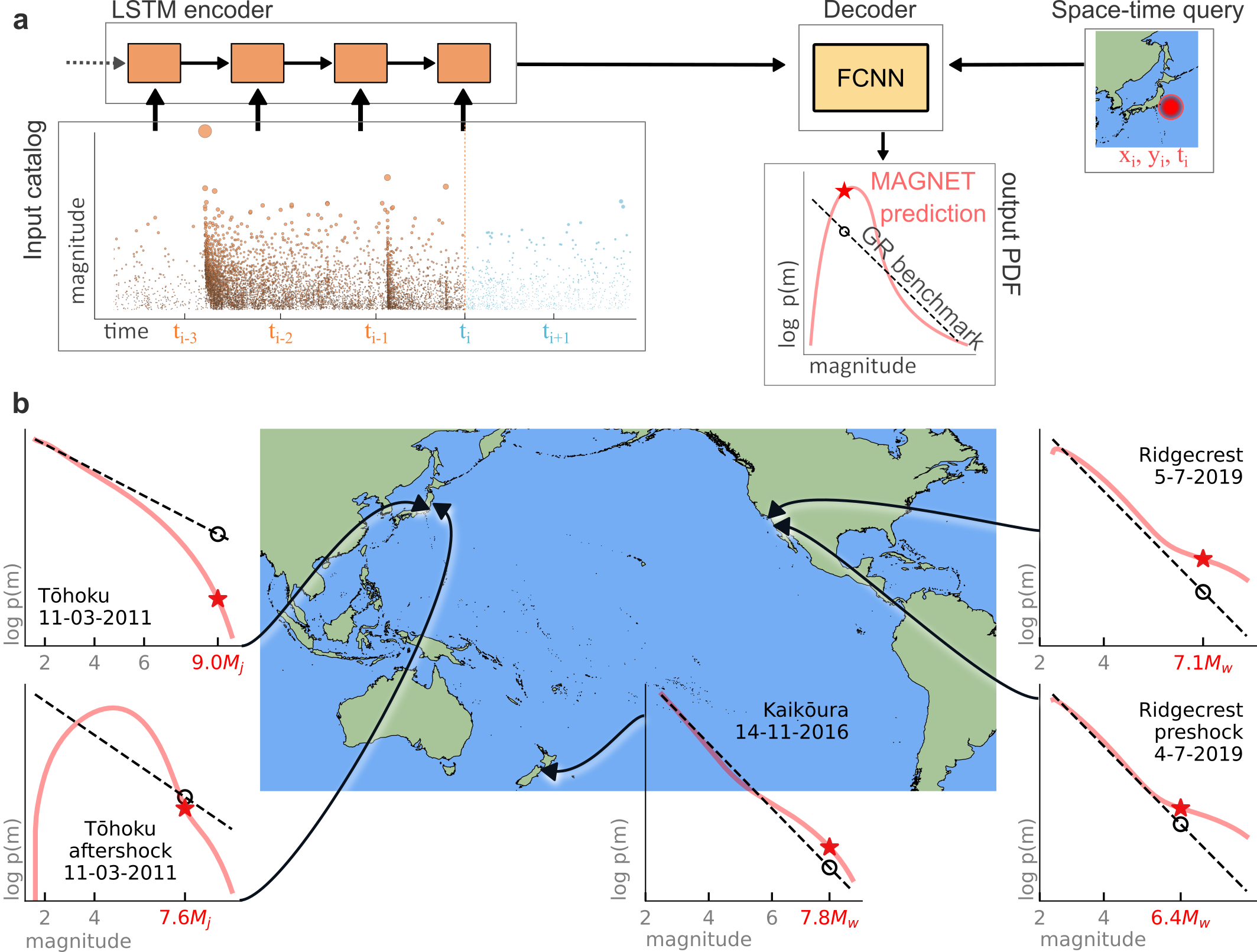}
    \caption{\textbf{a} MAGnitude Neural EsTimation model (MAGNET) task and architecture. Seismic history is continuously encoded using a Long Short-Term Memory neural network. For each earthquake, the encoded history up to its occurrence time ($t_i$) is concatenated with the time and location of the earthquake (``space~~time query''). The combined information is fed into a fully-connected neural network that outputs a probability density function (PDF) of the predicted magnitude at that time and location. In turn, the output is compared to a benchmark model (e.g. GR in dashed line) \textbf{b} Performance on major earthquakes: MAGNET's resulting PDFs for a selection of well-known major earthquakes worldwide from the test set. The red curve represents the model's predicted PDF and the dashed black curve represents the Gutenberg-Richter (GR) magnitude-frequency distribution, a common naive benchmark. For most earthquakes shown here, the likelihood of the observed magnitude (red star) is higher for MAGNET than for the GR benchmark (black circle). Notably, the T\=ohoku aftershock shows a qualitatively different behavior.
    }
    \label{fig:intro_fig}
\end{figure*}

\section{Methodology}
\subsection{Model architecture}
MAGNET is a neural network (NN) that takes as input a hypocentral catalog of regional seismicity, and the time and location of a future event. The NN produces a probability density function (PDF) estimating its magnitude.
Examples of the resulting PDFs for a few major earthquakes are presented in Fig. \ref{fig:intro_fig}.
These PDFs are not to be interpreted as an estimation of the instantaneous histogram of magnitudes (a ``frequentist'' interpretation) but rather should be interpreted in Bayesian terms: they are the model estimation of the magnitude of a specific future event.

To generate these predictions, MAGNET adopts an encoder-decoder architecture.
Here we briefly describe the main design; full details are provided in the Methods section and Supplementary Fig.~\ref{fig:architecture}.
At each time $t$, the encoder processes the seismic catalog up to this time by calculating a set of predefined statistics and characteristics (``features'') of regional seismicity. For example, the total moment release in the last week, the distance to the last earthquake above a certain magnitude, and so on. Such feature functions are common in other earthquake modeling approaches\cite{adeli_probabilistic_2009, panakkat_neural_2007, devries_deep_2018, asim_earthquake_2017, martinez-alvarez_determining_2013, Kong2019, xi_deep_2024, rouet-leduc_machine_2017-2, wan_advances_2024}.
MAGNET uses encoding functions, with varying time windows into the past, allowing the model to capture temporal patterns on scales between minutes and years.
The time sequence of these features is then fed into two recurrent neural network, specifically long-short term memory (LSTM) units~\cite{hochreiter_long_1997}. These LSTMs are coined by us the \textit{Recent Earthquakes Encoder} and the \textit{Seismicity Rate Encoder } to indicate their role in encoding the seismic history. The output of these two encoders is a representation of the system's state at time $t$.
The decoder takes as input the encoded seismic history and the space-time coordinates of the event, ${\bf{x}}_i$, $t_i$ (with $t_i>t$), termed a ``query'', and outputs the parameters of the PDF. Its architecture is a fully connected neural network (FCNN).

Importantly, while the encoders use the entire available seismic history as features, we only train the model with queries of space-time coordinates where earthquakes indeed occurred, and only events above a certain threshold magnitude are used as queries. We set this threshold to be the estimated completeness magnitude of the train set.

We parametrize the distribution as a mixture of two stretched Kumaraswamy distributions \cite{kumaraswamy_generalized_1980}, which is a 5-parameter family. These 5 parameters are the output of the neural model.
This parametrization allows a smooth interpolation between power-law decay (resembling Gutenberg-Richter distribution) and localized distributions concentrated around specific magnitudes, allowing the model to produce both ``ignorant'' GR-like predictions and confident predictions focused on particular magnitudes.
This versatility is demonstrated in Supplementary Fig.~\ref{fig:kumaraswamy_example}.
The choice of this distribution family is a design choice that influences model performance, and other choices are also possible. We denote the output of the model, the probability distribution of the magnitude of an event by $p_{\textbf{x}_i, t_i}(m)$ (Fig. \ref{fig:intro_fig}a).

\subsection{Optimization and benchmarking}
During training, the model optimizes the average log likelihood of the observed magnitudes over the training set,
\begin{align}
    \mathcal{L} & = -\mathbb{E}[\ell_i] \ ,
                &
    \ell_i      & =\log\left(p_{\textbf{x}_i, t_i}(m_i)\right)\ .
    \label{eq:likelihood}
\end{align}
where $\ell_i$ is the likelihood score of the $i$-th event, and the expectation is the empirical average over the training set.
Each dataset is divided into non-overlapping training, validation, and test time spans, see Supplementary Table \ref{tab:set_details} for details. Models were optimized on the training set, with hyperparameter tuning performed on the validation set.
All metrics reported in this work are evaluated solely over the test set, which was not used for model training or hyperparameter selection.

We benchmark the model against various stationary and quasi-stationary models.
The difference between MAGNET's likelihood score and that of a benchmark is called \textit{information gain} and quantifies the reduction in uncertainty about the magnitude that MAGNET produces over the benchmark.
By definition, if measurements are perfect and earthquakes magnitudes are independent of seismic history, our model cannot show any information gain on average.

However, as discussed in the introduction, comparison of likelihood scores to stationary distribution might produce spurious information gain due to detection limitations.
Specifically, small earthquakes immediately following a large earthquake may not be detected (STI). The space- and time-dependent detecability threshold is called \textit{completeness magnitude} and is denoted by $m_c(\textbf{x},t)$.
A model that learns the detection artifact will produce higher likelihood even if the underlying distribution is time independent.
To account for this effect, we present the likelihood score only for events above the $m_c$, which are properly conditioned,
\begin{equation}
    p \left( m \vert m_i > m_c(t) \right) = \frac{p_{\pmb{x}_i, t_i}(m)} {\int_{m_c(t)}^{\infty} p(m') dm'}
    \label{eq:conditioned_likelihood}
\end{equation}
This metric, assuming $m_c$ is evaluated correctly, cannot show information gain due to STI. We use various established methods to estimate $m_c$, as detailed in the methods section.

\section{Results} \label{sec:results}

We applied MAGNET on three distinct earthquake catalogs of diverse seismogenic regions: the Hauksson Catalog \cite{hauksson_waveform_2012} for Southern California, GeoNet \cite{gns_geonet_1970} for New Zealand, and the JMA catalog \cite{noauthor_japan_nodate} for Japan (set splits and regions are detailed in Supplementary Table~\ref{tab:set_details} and presented in Supplementary Fig.~\ref{fig:tested_regions}).
Since the catalogs for each regions are compiled using different measurement methodologies and exhibit varying data quality, a separate model is trained for each region.

Examples of the resulting PDFs are presented in Fig. \ref{fig:intro_fig}b for a few prominent earthquakes from the test set, superimposed on the stationary GR distribution (fitted on the train set), which is the naive benchmark.
It can be seen that MAGNET shows an information gain over the GR benchmark for most examples, and in some cases MAGNET's prediction differs qualitatively in terms of the distribution shape.
For instance, MAGNET shows an information gain of approximately 3.7 and 2 bits for the Ridgecrest mainshock in Southern California and the Kaik\=oura event in New Zealand, respectively, over the naive GR predictor.
Quantitative analysis in the results section will demonstrate that MAGNET shows consistent information gain over all benchmarks across all test regions, demonstrating that earthquake magnitudes are not history independent.

\begin{figure*}[ht!]
    \centering
    \includegraphics[width=1\textwidth]{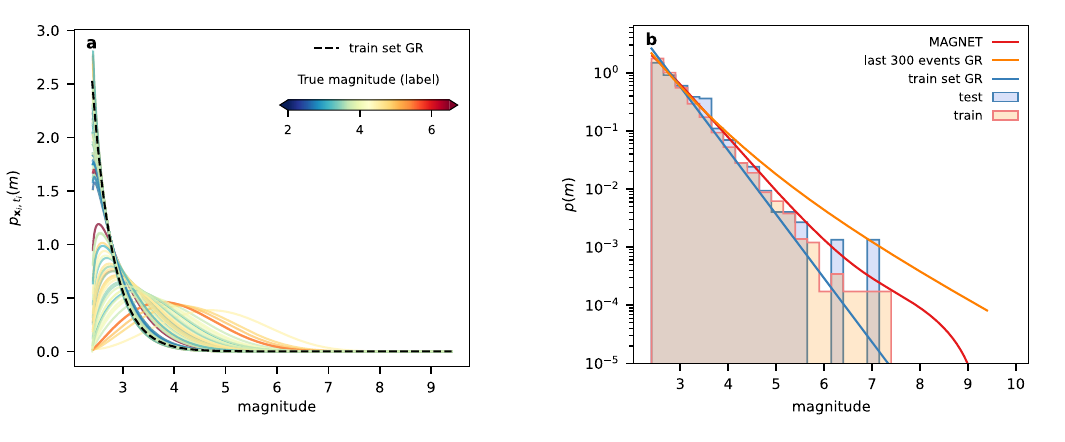}
    \caption{
        \textbf{MAGNET's output}. \textbf{a},
        Predicted magnitude PDFs produced by MAGNET for 100 randomly sampled events from the test set in the Southern California region.
        To counter the imbalance in the catalog, which is dominated by small events, we sampled events with higher probability for large magnitudes, yielding an approximately uniform magnitude distribution for demonstrative purposes.
        Curves are colored according to the magnitude of each earthquake event. It is seen that the PDFs for high magnitude events are generally skewed towards higher magnitudes, as expected from a predictive model.
        We superimpose the train set Gutenberg-Richter distribution (dashed black line) with $\beta$ value ($=2.53$) and completeness magnitude $m_c$ ($=2.4$) fitted on the train set.
        \textbf{b},
        Marginal magnitude distribution produced by MAGNET, averaging all PDFs from the test set. The marginal distribution closely resembles the GR distribution but follows the test set distribution more precisely.
        Note that one of the benchmarks (last 300 events GR) is a superposition of exponents, so it does not exhibit a linear trend on a log scale like the train set GR.
        Histograms of the train (orange) and test (blue) sets are shown.
    }
    \label{fig:model_output}
\end{figure*}

For demonstration, Fig. \ref{fig:model_output}a shows the output probability density functions (PDFs) predicted by MAGNET for 100 randomly selected events from the Southern California dataset.
The PDFs exhibit a clear qualitative trend: those calculated for higher-magnitude events (warmer colors) are skewed towards larger magnitudes compared to lower-magnitude events (cooler colors).
For comparison, we present the naive, history-independent predictor, the stationary GR distribution, overlaid on MAGNET's predictions.
A more quantitative analysis of this data is shown in Supplementary Fig.~\ref{fig:info_gain_per_bin}, demonstrating the information gain of MAGNET over GR as function of magnitude.

It is interesting to note that while the PDFs of the individual events may qualitatively differ from the GR distribution, the average PDF over all events does resemble the marginal GR distribution.
This is demonstrated in Fig.~\ref{fig:model_output}b, presents the marginal PDF of magnitudes for MAGNET's prediction, $P(m)=N^{-1}\sum_i p_{\textbf{x}_i, t_i}(m)$.
For reference, the marginal distributions of two additional common benchmark model (the stationary GR fitted on the train set and a 300-event moving window GR presented in Gulia \& Wiemer 2019\cite{gulia_real-time_2019}) are superimposed with MAGNET's.
It can be seen that, although MAGNET was not explicitly constrained to do so, its average prediction distribution aligns well with the ``uninformed'' GR distribution.
We also observe that MAGNET's average prediction aligns well with the \emph{test set} empirical distribution, rather than the training set, suggesting its ability to generalize beyond the training data.
This observation is more pronounced in the New Zealand and Japan data sets presented in Supplementary Fig.~\ref{fig:model_output_em}.
These are promising qualitative results, demonstrating that the model works as expected.

We now turn to a thorough quantitative assessment of the model performance, benchmarked against various stationary and quasi-stationary magnitude distributions and estimation methods of $m_c$.
The main metric underlying model comparison is the information gain for individual events,
\begin{equation}
    \Delta \ell_i = \log  p_{\pmb{x_i}, t_i}^{M} (m_i \vert m_i>m_c) - \log p^{B}(m_i \vert m_i>m_c)
    \label{eq:information_gain}
\end{equation}
where the superscript $M$ stands for MAGNET likelihood score, and $B$ for that of the benchmark. The conditioning on $m_c$ is done according to Eq. \ref{eq:conditioned_likelihood}.
To support our main point, that earthquake magnitudes are not independent on seismic history, we evaluate Eq.~\ref{eq:information_gain} for many benchmark models $B$ and estimations of $m_c$.

The distributions we benchmark against are stationary Gutenberg-Richter (GR) and stationary Kumaraswamy distributions fitted to the training set, a spatially varying GR distribution constructed by a method proposed by Taroni et.~al.~(2021)\cite{taroni_highdefinition_2021}, and a host of time-dependent distributions. These include GR distributions computed using moving windows of 300 and 500 most recent events, alongside additional GR variants computed over moving windows spanning from 10 to 1000 past days, kernel density estimation of the magnitude distributions for the past 50 and 300 events, and the best fit of a Kumaraswamy distribution over the past 300 events.
Details of the computational methods are provided in the Methods section. Quantitative results for all benchmarks are given in Supplementary Information.

Similarly, in the main text we use $m_c(t)$ calculated by the b-stability algorithm \cite{woessner_assessing_2005-1} to account for detectability variation artifacts, which is calculated over the current 300 events, similar to the technique used in Gulia et. al., 2018\cite{gulia_real-time_2019}.
The results demonstrate that MAGNET's positive information gain is robust to the choice of $m_c$ estimation method and magnitude predictor across all 3 regions studied in this work. Additional methods for calculating $m_c(t)$ include the b-stability algorithm over the past 500 and 1000 events, the maximum curvature method \cite{woessner_assessing_2005-1} over the past 300, 500 and 1000 events, and the closed form proposed by Helmstetter et.~al.\cite{helmstetter_comparison_2006} fitted per region.

\begin{figure*}[ht!]
    \centering
    \includegraphics[width=1\textwidth]{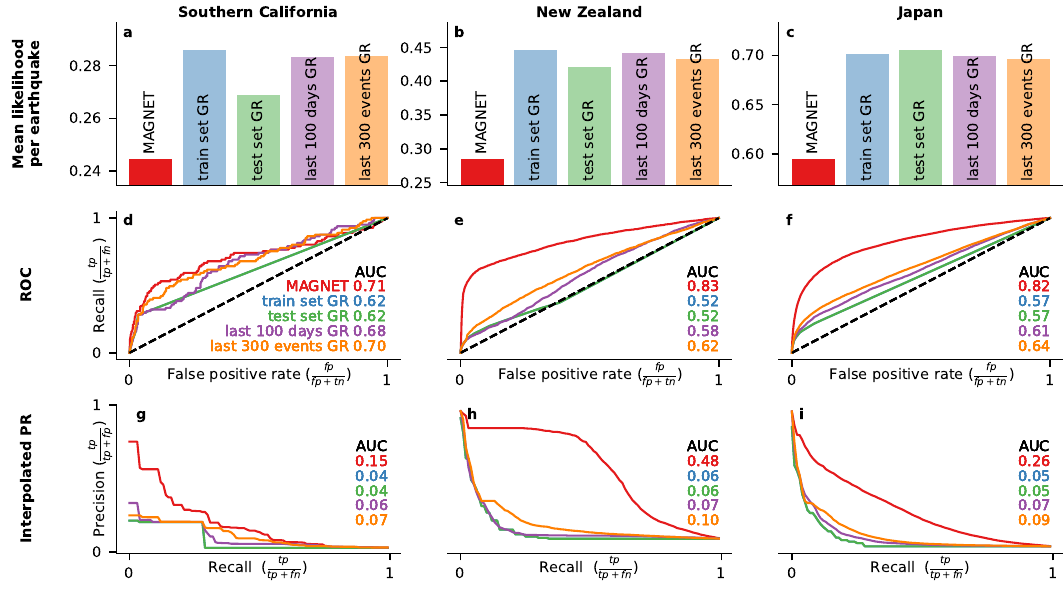}
    \caption{
        \textbf{Benchmarking MAGNET.} Comparison of MAGNET's temporally conditioned (eq. \ref{eq:conditioned_likelihood}) performance to common benchmarks. Non conditioned results are presented in Supplementary Fig.~\ref{fig:metrics_not_conditioned}. \textbf{a}, \textbf{b}, \textbf{c}, Minus mean information content of our MAGNET model (red) and other common benchmark magnitude predictors (see labels on bars in the figure).
        A lower score indicates a better preforming model.
        \textbf{d}, \textbf{e}, \textbf{f}, Receiver Operating Characteristic (ROC) and \textbf{g}, \textbf{h}, \textbf{i} the interpolated precision-recall (PR) curve for a binary classifier determining the next event will be large ($m\ge4$).
        The performance of such a classifier can be quantified by the area under the curve (AUC), noted in each frame, color coded identically to the bar plots.
        For the AUC metrics, a higher score indicates a better preforming classifier. Black dashed lines indicate the ROC curve of a random classifier.
    }
    \label{fig:metrics}
\end{figure*}

The main results are presented in Fig. \ref{fig:metrics}a-c, and the mean log-likelihood scores are presented in Table \ref{tab:mean_ll_benchmarks_main_text}.
Tables \ref{tab:mean_ll_all_benchmarks}, \ref{tab:mean_ll_additional_mc_cond}, \ref{tab:mean_ll_helm_conditioning} and Supplementary Fig.~\ref{fig:temp_incompleteness} show a comprehensive scan of all magnitude distribution benchmarks and $m_c(t)$ estimation methods.
Overall, these results indicate that MAGNET maintains an information gain regardless of the chosen $m_c$ estimation method or region with the sole exception of a spatially-variable GR distribution conditioned over a temporally-variable $m_c$ in New Zealand, which shows better but comparable metrics to MAGNET.

Another method to assess MAGNET's predictive power is to pose a binary question: will the magnitude of the earthquake at the provided space-time be higher than a given $m_t$?
The answer is easily calculated from output of the model $p_{\bf{x}_i, t_i}(m)$, since by definition:
\begin{equation}
    P(m_i>m_t)=\int_{m_t}^{\infty}p_{\bf{x}_i, t_i}(m)dm
    \label{eq:binary_question}
\end{equation}

In this setting, an event with $m_i > m_t$ is treated as a positive outcome (1), and an event with $m_i \leq m_t$ as a negative outcome (0).
This task is more interpretable and can be evaluated using standard binary classification metrics such as area under the receiver operator (ROC) curve \cite{Murphy} and under the precision-recall (PR) curve \cite{buttcher_information_2010}.
Figure~\ref{fig:metrics} show these metrics for all three regions with $m_t=4$ and compare against the GR-variant models. Again, MAGNET outperforms all benchmarks.

\begin{table*}[ht!]
    \centering
    \includegraphics[width=1\textwidth]{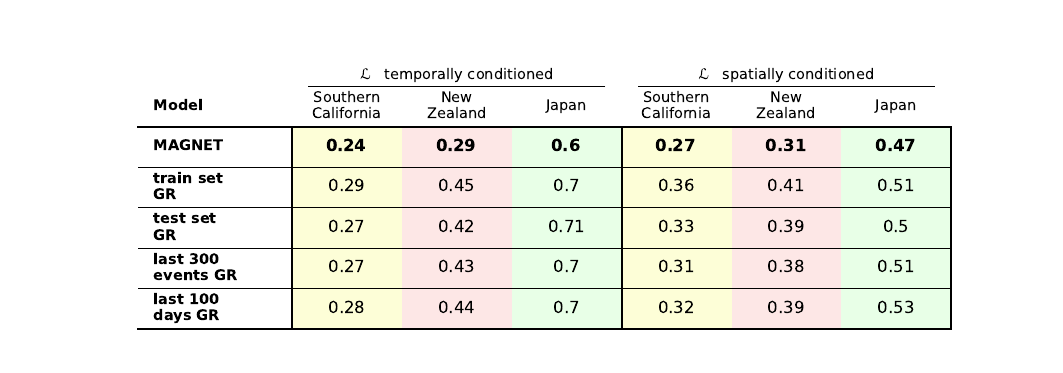}
    \caption{Mean score, $\mathcal{L}$, for various tested benchmarks, temporally and spatially conditioned.
        $\mathcal{L}$ is computed by Eq. \ref{eq:likelihood} and conditioned according to Eq. \ref{eq:conditioned_likelihood}. Lower score indicates a better magnitude predictor, best score in column is indicated in bold. First and second column triplets display the scores for the temporally and spatially conditioned $\mathcal{L}$ scores, respectively.
        The unconditioned $\mathcal{L}$ score is presented in Supplementary Table \ref{tab:mean_ll_all_benchmarks}, together with more magnitude predictors.
    }
    \label{tab:mean_ll_benchmarks_main_text}
\end{table*}

\begin{figure*}[ht!]
    \centering
    \includegraphics[width=1\textwidth]{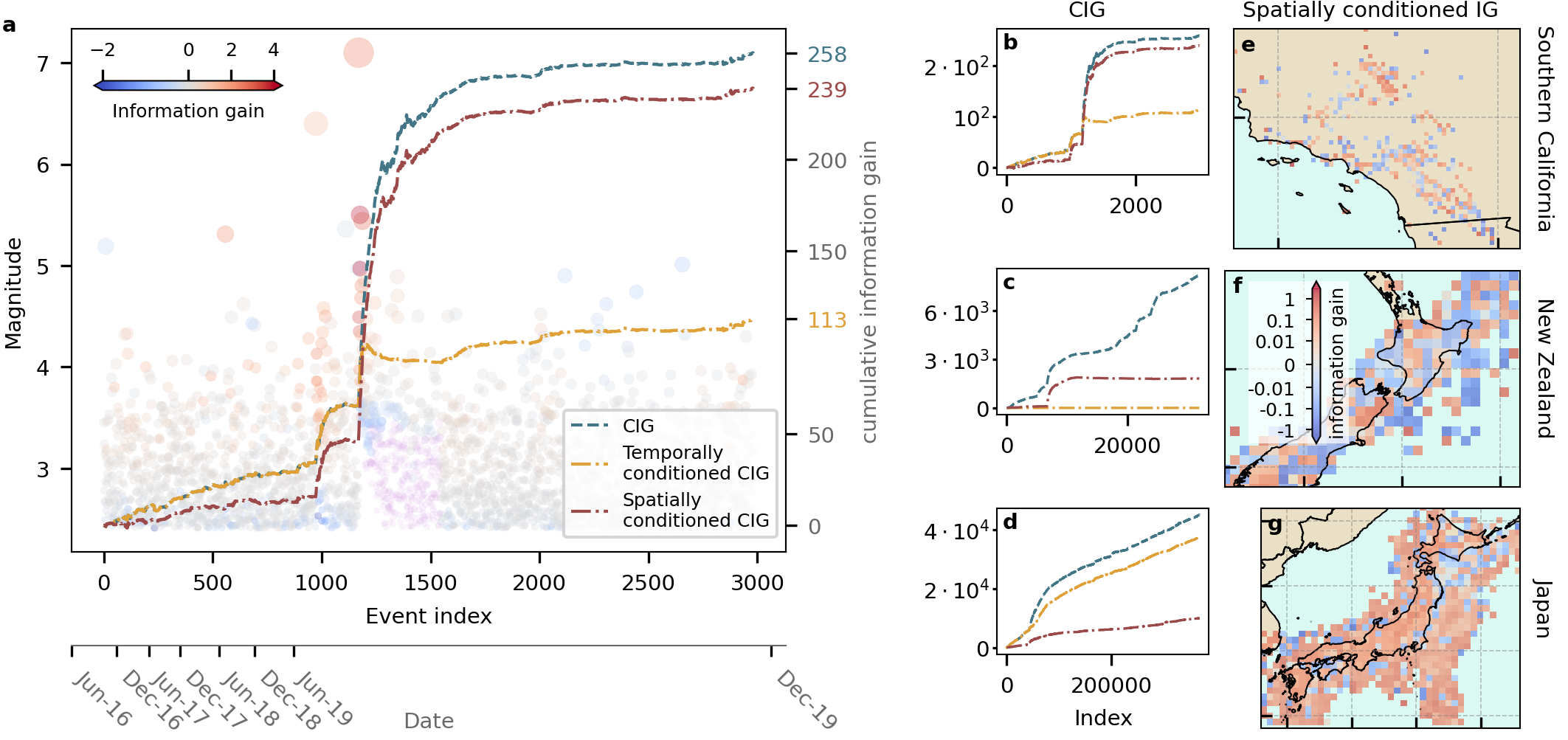}
    \caption{
        \textbf{Information gain in MAGNET. a},
        Information gain of individual events in the Southern California test set.
        Scattered dots indicate magnitude at event index, colored by the temporally conditioned information gain (IG) per event (scale is indicated by colorbar), where warm colors indicate an advantage for MAGNET, and cool colors indicate an advantage for the benchmark. The secondary horizontal axis (grey) indicates the corresponding origin time. Events below the temporal incompleteness, i.e. conditioned IG cannot be calculated, are plotted in purple.
        Cumulative Information gain (CIG) (dashed blue), temporally conditioned CIG (dashed yellow), and spatially conditioned CIG (dashed red) curves are superimposed on the scatter, demonstrating an increasing function for most times.
        \textbf{b-d},
        Information gain (dashed blue) and temporally and spatially conditioned CIG (dashed yellow and red respectively) curves for Southern California, New Zealand, and Japan test sets, respectively.
        An exception from the increasing trend can be observed for the spatially conditioned CIG of New Zealand (red curve in c).
        \textbf{e-g} Spatial distribution of spatially conditioned information gain for all regions in the study.
        Maps show the difference in mean information content per bin, with the model's advantage indicated by warm colors and disadvantage by cool colors, as defined by the colorbar adjacent to f. The full region is presented in Supplementary Fig.~\ref{fig:tested_regions}.
    }
    \label{fig:info_gain_over_time}
\end{figure*}

It is interesting to break down the total information gain of MAGNET, reported in Table \ref{tab:mean_ll_benchmarks_main_text}, to individual events and see for which events the model achieves its advantage.
For example, Fig. \ref{fig:info_gain_over_time}a shows a the time sequence of seismicity during the test period in Southern California, where events are colored according to their individual information gain, defined in Eq. \ref{eq:information_gain}.
In addition, the cumulative information gain (CIG) is plotted.
It is seen that CIG increases almost monotonically, suggesting that our model is consistently advantageous, and its information gain is not restricted to short intermittent periods, such as immediate aftershock sequences.
Noticeably, after the main shock of the Ridgecrest sequence, the CIG curves (Fig. \ref{fig:info_gain_over_time}, yellow) are decreasing, signaling a short-lived period of poor performance. This phenomenon is also observed following the New Zealand Kaik\=oura earthquake (7.8$M_W$, Fig. \ref{fig:info_gain_over_time}c, and Supplementary Fig.~\ref{fig:info_gain_EM}c,e).
A similar phenomenon is observed in the Japan test set, there MAGNET’s information gain for events of magnitude above $\sim 7$ following the 2011 Tōhoku earthquake are consistently negative.
This performance drop likely reflects a fundamental shift in the mechanical state and seismicity properties of the region, a change previously documented in the literature. \cite{kumazawa_quantitative_2013, terakawa_changes_2013}
However, these instances are rare, and the model shows consistent information gain during periods of both low and high seismicity. This is demonstrated in Supplementary Fig.~\ref{fig:info_gain_per_bin}d-i, which shows no clear correlation between seismicity rate and information gain.
Another interesting observation is that, perhaps surprisingly, the foreshock and main shock events demonstrate positive information gains over the GR benchmark, with $\Delta \ell$ values of approximately $1.8$ and $3.7$ bits, respectively. These events can be seen in Fig. \ref{fig:info_gain_over_time}a as the two highest magnitude warm-colored dots.

The corresponding figures for the other regions are presented in Fig. \ref{fig:info_gain_over_time}b-d and show qualitatively similar results.
Specifically, we show an average information gain of $\sim 0.15$ bits per earthquake in the same region in Japan where Ogata et.~al.~\cite{ogata_exploring_2018} could not find an information gain using spatiotemporal GR variations. Similar plots for the other studied regions are shown in Supplementary Fig.~\ref{fig:info_gain_EM}.

A spurious information gain may also be caused by temporally constant but spatially varying earthquake detection thresholds, due to the structure of the seismic network \cite{schorlemmer_probability_2008, mignan_bayesian_2011}. This implies that the local completeness magnitude is also generally spatially dependent. In such a case, the information gain of MAGNET might not imply that the magnitudes are correlated in time.
To examine this, we perform an additional test similar to the temporal analysis. We calculate the information gain conditioned on the spatially-varying completeness magnitude. Specifically, we replace the temporal incompleteness function, $m_c(t)$, in Eq. \ref{eq:conditioned_likelihood} with a spatial varying variant $m_c(x,y)$ estimated for a coordinate grid of $0.1^\circ$. The resulting $m_c(x,y)$ distribution is presented in Supplementary Fig.~\ref{fig:mc_maps}, further details about the computation are given in the Methods section.

The spatially conditioned CIG are shown by the red dashed curves in Fig. \ref{fig:info_gain_over_time}. It is a steadily increasing function for the Southern California and Japan data sets.
In the New Zealand data set the curve shows an initial increase followed by a plateau after the first sequence of major events (the 2016 Te Araroa and Kaik\=oura earthquakes).
This may indicate that a good portion of this information gain followed by these earthquakes does originate from spatial inhomogeneity, as also hinted by the spatially-varying GR benchmark.
Nevertheless, the overall score still shows an advantage for our model over the test set.
Figure \ref{fig:info_gain_over_time}e-g shows the spatial distribution of MAGNET's advantage over the GR benchmark. All three regions exhibit no clear spatial pattern, suggesting that information is not gained solely from any one specific area (compare Supplementary Fig.~\ref{fig:seismicity_grid} for the distribution of seismicity). Similarly, one can decompose the information gain to magnitude bins and ask whether the model tends to perform well on high- or low-magnitude events.
We perform this analysis in Supplementary Fig.~\ref{fig:info_gain_per_bin}, which shows a slight advantage at medium and large earthquakes.

\section{Implications and Outlook}
In this work we have used a neural model to find a statistical dependence between earthquake magnitudes and seismic history. Since our model is provided with the known location and timing of an event and is only tasked with predicting its magnitude, we have separated out the question of rate and location of earthquakes, and directly probe whether their magnitude is history dependent. MAGNET shows a clear and robust information gain against a host of benchmarks, after accounting for possible measurement artifacts. Our results indicate that the regional seismic history in the form of a hypo-center catalog contains information about the magnitude of a specific future earthquake.

This means that the seismic patterns preceding large events are distinguishable, at least to some extent, from those preceding ``background'' activity.
A similar claim was recently raised by Ben-Zion and Zaliapin\cite{ben-zion_localization_2020}, which show trends in statistical properties of the seismicity prior to fault-size events, though a direct linkage between the two observation requires further investigation.
It would also be interesting to understand what are the statistical features of seismicity that our model identifies.
Specifically, future work should include ablation studies to determine the
relative informativeness of the individual encoders, e.g. recent history versus local seismicity rate, to provide deeper insights into the physical mechanisms the model is leveraging.
This will be the subject of future research dedicated to the interpretability of the MAGNET architecture.

Understanding such patterns will be significant in the effort of more general earthquake prediction~\cite{mousavi_deep-learning_2022, mousavi_machine_2023, mignan_neural_2020, karimpouli_explainable_2023, bergen_machine_2019}, though this is still further down the road.
Additionally, such models may be incorporated in early warning systems, which already provide a quick (though not accurate) estimation of the source time and location.

Another possible application is incorporating a history-dependent magnitude distribution into point-process rate forecasting models, such as the epidemic type aftershock sequence model (ETAS) or similar \cite{ogata_statistical_1988, ogata_statistics_2017, dascher-cousineau_using_2023}, which currently assume no statistical dependence of the magnitude.
Since larger earthquakes produce more aftershocks \cite{kagan_earthquakes_2014}, a more accurate estimation of magnitudes will have a positive compounding effect on the resulting statistics.
Furthermore, utilizing ETAS-type model as a benchmark in future studies would provide a measurable improvement in rate-forecasting and further reassurance that MAGNET effectively extracts meaningful information from seismic catalogs.

\FloatBarrier
\clearpage

\section{Methods}
\subsection{Neural architecture}
A detailed visual of the model's architecture is presented in Supplementary Fig.~\ref{fig:architecture}. The catalog is used as an input into three distinct encoding components:

\textbf{\textit{Recent Earthquakes Encoder.}}
This encoder calculates $N_{features}$ predefined (non-trainable) functions over the $N_{max} = 80$ most recent events above the train set's calculated $m_c$ within a lookback period of 7 years.
The list of predefined functions is provided in Supplementary Table \ref{tab:recent_earthquake_features}.
These features are calculated per each of the $N_{max}$ events and stored in a tensor of shape $[N_{max},\; N_{features}]$.
Such tensor is calculated for each query timestamp and in turn fed into to a Long-Short-Term-Memory Neural Network (LSTM NN).
The structure of this encoder is visually presented in Supplementary Fig.~\ref{fig:architecture}b.

\textbf{\textit{Seismicity Rate Encoder.}}
This encoder estimates the amount of energy released in a given region around the query coordinates over various past time windows.
For each threshold magnitude $m_t \in \{2, 3, 4, 5, 6\}$, it calculates the sum $\sum_{i}e^{m_i}$ for all events above $m_t$ within the time periods $\left[ t-b_k, t-b_{k-1} \right]$, where $b_k$ is taken from the ordered set $\left( 1 \text{hr}, 12 \text{hr}, 2 \text{d}, 10 \text{d}, 100 \text{d}, 3 \text{yr}, 10 \text{yr}, 30 \text{yr} \right)$.
The surrounding region is defined as a cell with a side length of $0.5^{\circ}$.
The results of these calculations are stores in a tensor of shape $[N_{m_t},\; N_{b}]$, where $N_{m_t}$ is the number of threshold magnitudes and $N_b$ is the number of time windows.
This tensor is calculated for each query timestamp and in turn fed into to a LSTM NN.
The structure of this encoder is visually presented in Supplementary Fig.~\ref{fig:architecture}c.

\textbf{\textit{Space-time Query Encoder.}} The encoder directly passes the spatial query coordinates and the time difference from the most recent past event, in seconds. No further processing is done in this encoder.

\textbf{\textit{Fully Connected Neural Network Decoder.}} Outputs of all encoders are concatenated and fed to a FCNN (Decoder). The sizes of layers in the FCNN, as the number of layers themselves, together with other hyperparameters are optimized by cross-validation: the hyperparameters were scanned over a predefined grid, and the configuration yielding the best performance on the validation set was selected. The resulting values are displayed in Supplementary Table \ref{tab:hyperparameters}.
These hyperparameters and other more architecture specific are set as the default values in the configuration files (\textit{*.gin} files) in the open source repository of MAGNET.
A $\texttt{tanh}$ activation is used for the hidden layers, and a $\texttt{softplus}$ as an activation on the output layer. Training process uses the Adam optimizer\cite{kingma_adam_2017}.

\begingroup
\setlength{\dblfloatsep}{2\baselineskip}
\setlength{\floatsep}{2\baselineskip}

\begin{table*}[htbp]
    \centering
    \begin{tabular}{c|c|c|c}
        \multicolumn{4}{c}{Selected hyperparameters of neural architecture and training} \\
                                   & Southern California & New Zealand   & Japan         \\
        \hline
        Decoder (FCNN) layer sizes & [256, 64, 16]       & [256, 64, 16] & [256, 64, 16] \\
        Learning rate              & 0.0001              & 0.0001        & 0.001         \\
        Batch size                 & 256                 & 64            & 64            \\
        Training epochs            & 150                 & 150           & 50            \\
    \end{tabular}
    \caption{Selected hyperparameters of neural architecture and training.}
    \label{tab:hyperparameters}
\end{table*}

\subsection{Loss metric}
To obtain the Probability Density Function (PDF) of the magnitudes ($p(m)$) for each query coordinates ($\textbf{x}_i, t_i$) we maximize the likelihood, $\mathcal{L} = -\langle \ell_i \rangle$, where $\langle \cdot\rangle$ is the mean over all examples in the set. The PDF is taken to be a mixture of two stretched and shifted Kumaraswamy distributions~\cite{kumaraswamy_generalized_1980}:
\begin{widetext}
    \begin{equation}
        p\left( m \right)
        \equiv
        \sum_{j=1,2} \frac{A_j}{\sigma}a_jb_j\left(\frac{m-m_c}{\sigma}\right)^{\left(a_j-1\right)}\left(1-\left(\frac{m-m_c}{\sigma}\right)^{a_j}\right)^{\left(b_j-1\right)}
        \label{eq:kumaraswamy}
    \end{equation}
\end{widetext}

Here $a_j, \ b_j$ are the parameters of the Kumaraswamy distribution.
Several realizations of a single Kumaraswamy distribution with varying $a,b$ parameters are presented in Supplementary Fig.~\ref{fig:kumaraswamy_example}.
$m_c$ is the training set's completeness magnitude. $m_c$ together with $\sigma$ define the support of $p(m)$ to $[m_c, m_c+\sigma]$. The specific parameters chosen are detailed in Supplementary Table \ref{tab:pdf_support_parameters}.
$A_j$ is a normalization prefactor, with $j$ the summation index defining the PDF mixture.

\begin{table*}[htbp]
    \centering
    \begin{tabular}{c|c|c|c}
        \multicolumn{4}{c}{PDF support parameters}                   \\
                 & Southern California         & New Zealand & Japan \\
        \hline
        $m_c$    & 2.4                         & 2.5         & 1.6   \\
        \makecell{$m_c$                                              \\calculation method}
                 & \makecell{Maximal curvature                       \\+ 0.2 margin}
                 & \makecell{Maximal curvature                       \\+ 0.2 margin}
                 & \makecell{b-stability}                            \\
        $\sigma$ & 7                           & 7           & 9     \\
    \end{tabular}
    \caption{PDF support parameters.}
    \label{tab:pdf_support_parameters}
\end{table*}
\endgroup

\subsection{Definition of benchmark models}
\textit{Last $n$ events} The Gutenberg- Richter (GR) distribution fitted for the past $n$ events. This method follows Gulia \& Wiemer\cite{gulia_real-time_2019} for constructing a b-value time-series. If in the last $n$ events there are less than 50 earthquakes of magnitudes above $m_c^{train}$, then additional events from the past are included, until this condition is met. $m_c^{train}$ is estimated using the maximal curvature method\cite{wiemer_minimum_2000}.

\textit{Last $d$ days} The Gutenberg- Richter (GR) distribution fitted for the events in the past $d$ days. This method is identical to the previous method in all but the definition of the window selection. Here, as in the \textit{Last $n$ events} method, if there are less than 50 earthquakes of magnitudes above $m_c^{train}$ in the past $d$ days, then additional events from the past are included, until this condition is met.

\textit{Spatially varying GR} Estimation of the local $b$ value is done following the method of Taroni et.~al.~\cite{taroni_highdefinition_2021_EM}. Seismicity data is
binned into $0.1^\circ \times 0.1^\circ$ bins, and the GR distribution is fitted for each bin. $b$ for a specified location $x_i$ is calculated by the average of all $b$ values in bins within 30km around $x_i$. Bins which contain less than 150 events are discarded. Therefore, this benchmark is evaluated on a smaller subset of events compared to other benchmarks and MAGNET and is marked by an asterisk (*) in the results Supplementary Tables \ref{tab:mean_ll_all_benchmarks}, \ref{tab:mean_ll_additional_mc_cond} and \ref{tab:mean_ll_helm_conditioning}. The comparison over an identical set of events is presented in Supplementary Table~\ref{tab:mean_ll_diff_MAGNET_spatial_GR}.

\textit{Kernel density estimation (KDE) of last n events} Selecting the last $n$ events (similar to described in \textit{last $n$ events}) we compute the KDE estimation of the magnitude PDF, using the 'scott' estimator bandwidth\cite{scott_2015}. This is implemented using the SciPy tool\cite{scipy_2020}. The resulting PDF is used as the prediction of the magnitude.

\textit{Kumaraswamy PDF fit of last n events} Selecting the last $n$ events (similar to described in \textit{last $n$ events}) we compute the best fit of a mixture of two stationary Kumaraswamy distributions (Eq. \ref{eq:kumaraswamy}) to the magnitude histogram of the selected events, using an expectation-maximization algorithm.
The resulting PDF is used as the prediction of the magnitude.

\subsection{Measurement of incompleteness}

\textit{\textbf{Data set incompleteness}} The incompleteness of the entire train set, $m_c$ is calculated per region. This is used as a threshold above which the model's prediction is evaluated (lower magnitude events are used as features, but metrics are not reported for them). $m_c$ is calculated using the maximal curvature method\cite{wiemer_minimum_2000} with a safety margin of $0.2$ units for the Southern California and New Zealand data sets, and the b-stability method\cite{woessner_assessing_2005-1} for the Japan data set.

\textit{\textbf{Temporal incompleteness}} For conditioning on temporal incompleteness, ee estimate $m_c(t)$ using multiple methods and compare the results.
Temporal incompleteness is measured at the times of events by fitting the completeness magnitude to a window of 300 (or 500) events: 150 (250) past events, 1 (1) current event, 149 (249) future events. The window is constructed by the same algorithm as in the \textit{Last $n$ events benchmark}.
In the main text we use the \textit{b-stability} method \cite{woessner_assessing_2005-1}.

\textit{b-stability method} \cite{woessner_assessing_2005-1} is considered a reliable method for estimating the completeness magnitude. In the main text, the results presented are conditioned on this method of calculation for a window of 300 events. The results for conditioning on the b-stability method for windows of 500 and 1000 events are presented in Supplementary Table \ref{tab:mean_ll_additional_mc_cond}.

\textit{Maximal curvature method} \cite{wiemer_minimum_2000} is used with a margin of $0.2$ units. The results for conditioning on this computation for windows of 300, 500 and 1000 events are presented in Supplementary Table \ref{tab:mean_ll_additional_mc_cond}.

\textit{Helmstetter formula}\cite{helmstetter_comparison_2006} for estimation of short-term incompleteness

\begin{equation}
    m_c(t) = \max\{ m_i-G-H\log_{10}(t-t_i), m_{min} \}
    \label{eq:helmstetter}
\end{equation}

where $m_i$ is the magnitude of the main shock, to be considered above a specific threshold $m_t$, $G$ and $H$ are fitting parameters and $t-t_i$ is the time since the mainshock, in days.
The results for conditioning on this computation are presented in Supplementary Table \ref{tab:mean_ll_helm_conditioning}.
Here, we fit appropriate parameters ($ \left( m_t, G, H, m_{min} \right)$) for each catalog.
We find that the original parameters suggested in Helmstetter et    .~al.~2006\cite{helmstetter_comparison_2006} are the best fit for the Hauksson catalog of Southern California.
For the New Zealand and Japan data sets, we find that different sets of parameters may fit at different times, likely due to the large geographical areas covered by these catalogs.
Supplementary Table \ref{tab:mean_ll_helm_conditioning} presents the results for the best-fitting parameters for each catalog, including all relevant parameter sets for New Zealand and Japan.

\textit{Shifted Helmstetter formula} follows the same computational procedure as the original Helmstetter formula, with an additional shift applied to the final result to introduce a more conservative margin.

\textit{\textbf{Spatial incompleteness}} Spatial incompleteness is calculated for a grid of coordinates, spaced by $\theta^\circ$ degrees, covering the relevant region. $\theta$ is set to 0.1 for Southern California and 0.5 for New Zealand and Japan data sets. Calculation is done using train set's events. The value per point is calculated using the maximal curvature method\cite{wiemer_minimum_2000}, over the nearest, at least 100 events, with a minimal radius of $\theta^\circ$. In cases where the the nearest 100 events span over a radius of $2^\circ$ the data point is considered invalid. Events from the test set are divided into bins of $\theta^\circ$ centered around the point for which $m_c(x,y)$ was computed. Events are assigned with the $m_c$ value of their bin.
The resulting spatial completeness is presented in Supplementary Fig.~\ref{fig:mc_maps}.

\section*{Data and code availability}
The datasets analyzed during the current study are available in the relevant references mentioned throughout the article.
The code for analyzing the data is available in \url{https://github.com/Ner-Ber/MAGNET}. This is accompanied by the trained features and models found in the Zenodo repository \url{https://zenodo.org/records/14066185}.
Country borders are plotted using data from \url{https://geojson-maps.kyd.au/} .

\begin{acknowledgments}
    We thank Yehuda Ben-Zion, Assaf Inbal, Kelian Dascher-Cousineau, David Marsan, Eugenio Lippiello, Jiancang Zhuang and Yosihiko Ogata. for thoughtful discussions. YBS is supported by ISF grant 1907/22 and by Google Gift grant.
\end{acknowledgments}

\section*{Author contributions statement}
N.B., O.Z., O.G., Y.M. and Y.B.S. designed the research. N.B. conducted the experiments. N.B., O.G. and Y.B.S. preformed the statistical analysis. N.B. and O.Z. wrote the code base. N.B. and Y.B.S. wrote the manuscript. All authors reviewed the manuscript.

\section*{Additional information}
\textbf{Competing interests:} The authors declare no competing interests.

Please contact Yohai Bar-Sinai \href{mailto:ybarsinai@gmail.com}{ybarsinai@gmail.com} or Neri Berman \href{mailto:neriberman@gmail.com}{neriberman@gmail.com} for correspondence and requests, including questions regarding reprints and permissions.

\clearpage
\onecolumngrid
\makeatletter
\def\href@noop#1#2{#2}%
\def\selectlanguage#1{}%
\makeatother
\begingroup
\makeatletter
\@makeother\_
\makeatother
\small
\emergencystretch=2.5em
\bibliography{Magnitude_prediction_paper}
\endgroup
\clearpage

\renewcommand{\figurename}{Supplementary Figure}
\renewcommand{\tablename}{Supplementary Table}
\renewcommand{\thefigure}{S\arabic{figure}}
\renewcommand{\thetable}{S\arabic{table}}
\setcounter{figure}{0}
\setcounter{table}{0}
\renewcommand{\theHfigure}{S\arabic{figure}}
\renewcommand{\theHtable}{S\arabic{table}}

\begin{center}
    \LARGE\textbf{Supplementary Information for:}\\[0.5cm]
    \Large\textbf{Earthquake magnitudes depend on seismic history, as revealed by a neural network analysis}\\[1.5cm]

    \large
    Neri Berman$^{1,2,\ast}$, Oleg Zlydenko$^{2}$, Oren Gilon$^{2}$, Yossi Matias$^{2}$, and Yohai Bar-Sinai$^{1,3,4,\ast}$\\[1cm]

    \normalsize
    $^{1}$ Department of Physics, Tel-Aviv University, Tel-Aviv, Israel\\
    $^{2}$ Google Research, Google, Tel-Aviv, Israel\\
    $^{3}$ Racah Institute of Physics, The Hebrew University of Jerusalem, Jerusalem, Israel\\
    $^{4}$ The Rachel and Selim Benin School of Computer Science and Engineering, The Hebrew University of Jerusalem, Jerusalem, Israel\\[1cm]

    $^\ast$Corresponding authors: \url{ybarsinai@gmail.com}, \url{neriberman@gmail.com}
\end{center}

\vspace{1cm}

\noindent\textbf{This PDF file includes} supplementary figures and tables.

\vspace{0.5cm}

\noindent\textbf{Other supplementary materials for this manuscript include the following:}
\begin{itemize}
    \item Code for analyzing the data: \url{https://github.com/Ner-Ber/MAGNET}
    \item Trained features and models: \url{https://zenodo.org/records/14066185}
\end{itemize}

\clearpage
\begin{figure}[!tbp]
    \centering
    \includegraphics[width=\linewidth,height=0.6\textheight,keepaspectratio]{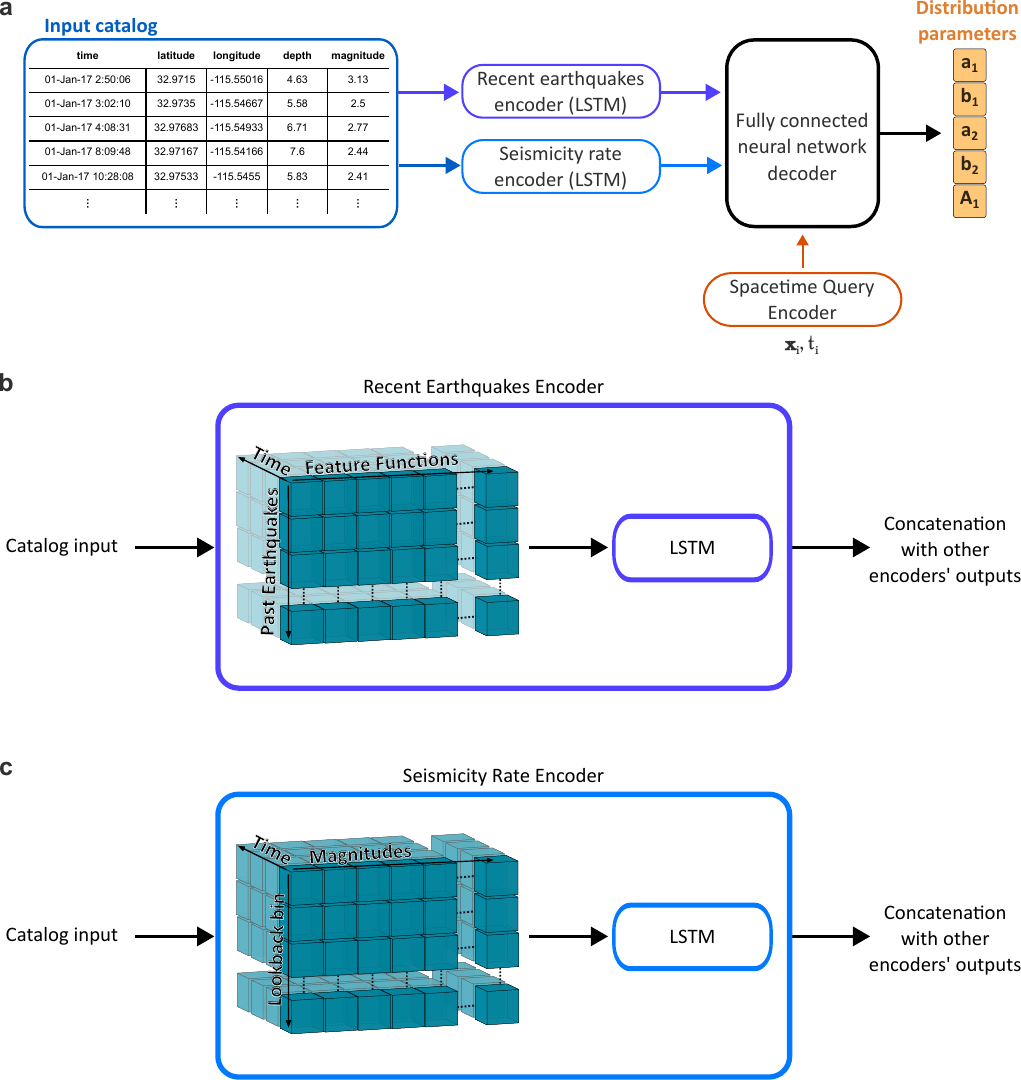}
    \vspace{\baselineskip}
    \caption{\textbf{MAGNET architecture.}
        \textbf{a}, Overall architecture of MAGNET. The catalog is processed by two encoders: the Recent Earthquakes Encoder (b), the Seismicity Rate Encoder (c). The resulting latent representation of these encoders is accompanied by the space-time query coordinates given by the Space-time Query Encoder.
        The outputs of all encoders are concatenated and passed to a fully connected neural network (FCNN) decoder, which outputs the parameters of the magnitude PDF.
        \textbf{b}, Detailed structure of the Recent Earthquakes Encoder. Predefined features are calculated for the $N_{max}$ most recent events above $m_c$ within a lookback period of 7 years.
        The resulting tensor is passed to a LSTM NN.
        \textbf{c}, Detailed structure of the Seismicity Rate Encoder. For several magnitude thresholds and nested time windows we compute a proxy for released seismic energy (\(\sum_i e^{m_i}\)) from events inside a $0.5^{\circ}$ cell around the query location for each time window. These measurements form a tensor per query timestamp that is fed into an LSTM.
        Architectures are described in detail in the Methods section.}
    \label{fig:architecture}
\end{figure}
\FloatBarrier

\begin{table}[htbp]
    \centering
    \begingroup
    \footnotesize
    \setlength{\tabcolsep}{3pt}%
    \renewcommand{\arraystretch}{0.92}%
    \setlength{\aboverulesep}{0.2ex}%
    \setlength{\belowrulesep}{0.35ex}%

    \begin{tabular*}{\textwidth}{@{}>{\raggedright\arraybackslash}p{45mm}>{\raggedright\arraybackslash}p{\dimexpr(\textwidth-45mm-6\tabcolsep)/3\relax}>{\raggedright\arraybackslash}p{\dimexpr(\textwidth-45mm-6\tabcolsep)/3\relax}>{\raggedright\arraybackslash}p{\dimexpr(\textwidth-45mm-6\tabcolsep)/3\relax}@{}}
        \toprule
        \textbf{Split}          & \textbf{Train}                                                                                                                                      & \textbf{Validation}                                                                     & \textbf{Test}                                                    \\
        \midrule

        \multicolumn{4}{@{}l}{\textbf{Southern California}}                                                                                                                                                                                                                                                                                        \\
        \midrule
        Time span               & [01-01-1981 00:00:00, \newline 03-02-2009 12:39:39)                                                                                                 & [03-02-2009 12:39:39, \newline 23-05-2016 23:05:46)                                     & [23-05-2016 23:05:46, \newline 31-12-2019 22:48:20)              \\
        Longitude span          & \multicolumn{3}{c}{Full catalog extent}                                                                                                                                                                                                                                                                          \\
        Latitude span           & \multicolumn{3}{c}{Full catalog extent}                                                                                                                                                                                                                                                                          \\
        Number of events        & 23,235                                                                                                                                              & 4,755                                                                                   & 2,983                                                            \\
        $N_{events} < 4$        & 22,647                                                                                                                                              & 4,547                                                                                   & 2,902                                                            \\
        $4 \leq N_{events} < 5$ & 526                                                                                                                                                 & 198                                                                                     & 73                                                               \\
        $5 \leq N_{events} < 6$ & 55                                                                                                                                                  & 9                                                                                       & 6                                                                \\
        $6 \leq N_{events} < 7$ & 5                                                                                                                                                   & 0                                                                                       & 1                                                                \\
        $7 \leq N_{events}$     & 2                                                                                                                                                   & 1                                                                                       & 1                                                                \\
        Notable earthquakes     & 1986 N. Palm Springs \newline 1992 Joshua Tree \newline 1992 Big Bear \newline 1992 Landers \newline 1999 Hector Mine                               & 2010 Baja California                                                                    & 2019 Ridgecrest                                                  \\
        \midrule

        \multicolumn{4}{@{}l}{\textbf{New Zealand}}                                                                                                                                                                                                                                                                                                \\
        \midrule
        Time span               & [05-01-1989 10:40:00, \newline 10-01-2008 21:20:00)                                                                                                 & [10-01-2008 21:20:00, \newline 13-05-2014 16:53:20)                                     & [13-05-2014 16:53:20, \newline 17-11-2022 07:42:07)              \\
        Longitude span          & \multicolumn{3}{c}{Full catalog extent}                                                                                                                                                                                                                                                                          \\
        Latitude span           & \multicolumn{3}{c}{(-50, -18)}                                                                                                                                                                                                                                                                                   \\
        Number of events        & 105,131                                                                                                                                             & 39,017                                                                                  & 32,055                                                           \\
        $N_{events} < 4$        & 99,759                                                                                                                                              & 37,254                                                                                  & 28,941                                                           \\
        $4 \leq N_{events} < 5$ & 4,993                                                                                                                                               & 1,610                                                                                   & 2,642                                                            \\
        $5 \leq N_{events} < 6$ & 350                                                                                                                                                 & 135                                                                                     & 449                                                              \\
        $6 \leq N_{events} < 7$ & 23                                                                                                                                                  & 12                                                                                      & 19                                                               \\
        $7 \leq N_{events}$     & 6                                                                                                                                                   & 6                                                                                       & 4                                                                \\
        Notable earthquakes     & 1989 Doubtful Sound \newline 1990 Weber \newline 1995 Offshore E. Cape \newline 2003 Fiordland \newline 2004 Puysegur Trench \newline 2007 Gisborne & 2009 Dusky Sound \newline 2010 Canterbury \newline 2011 Christchurch                    & 2016 Christchurch \newline 2016 Te Araroa \newline 2016 Kaikōura \\
        \midrule

        \multicolumn{4}{@{}l}{\textbf{Japan}}                                                                                                                                                                                                                                                                                                      \\
        \midrule
        Time span               & [01-01-1965 00:00:00, \newline 01-01-1999 00:00:00)                                                                                                 & [01-01-1999 00:00:00, \newline 01-01-2009 00:00:00)                                     & [01-01-2009 00:00:00, \newline 31-12-2017 23:59:59)              \\
        Longitude span          & \multicolumn{3}{c}{(125, 160)}                                                                                                                                                                                                                                                                                   \\
        Latitude span           & \multicolumn{3}{c}{(29, 49)}                                                                                                                                                                                                                                                                                     \\
        Number of events        & 206,374                                                                                                                                             & 243,783                                                                                 & 367,977                                                          \\
        $N_{events} < 4$        & 186,897                                                                                                                                             & 236,087                                                                                 & 356,542                                                          \\
        $4 \leq N_{events} < 5$ & 15,917                                                                                                                                              & 6,677                                                                                   & 10,049                                                           \\
        $5 \leq N_{events} < 6$ & 3,259                                                                                                                                               & 906                                                                                     & 1,207                                                            \\
        $6 \leq N_{events} < 7$ & 273                                                                                                                                                 & 98                                                                                      & 165                                                              \\
        $7 \leq N_{events}$     & 28                                                                                                                                                  & 15                                                                                      & 14                                                               \\
        Notable earthquakes     & 1968 Hyūga-nada \newline 1968 Tokachi \newline 1973 Nemuro \newline 1978 Miyagi \newline 1993 Hokkaidō \newline 1995 Great Hanshin                  & 2003 Tokachi \newline 2006 Kuril Islands \newline 2007 Kuril Islands \newline 2007 Noto & 2011 Tōhoku \newline 2011 Fukushima \newline 2012 Sanriku        \\
        \bottomrule
    \end{tabular*}
    \endgroup

    \footnotesize
    \vspace{\baselineskip}
    \caption{\textbf{Details of the datasets used in this study}. Each dataset is divided into training, validation, and test subsets. For each subset, the table reports the corresponding time span, the number of recorded events (including a breakdown by magnitude range), and examples of notable earthquakes. The splits are chronologically ordered and mutually exclusive to prevent information leakage between training, validation, and testing. Time spans use the notation '[' to indicate inclusion of the start time and ')' for exclusion of the end time.}
    \label{tab:set_details}
\end{table}
\FloatBarrier

\begin{figure}[ht!]
    \centering
    \includegraphics[width=1\textwidth]{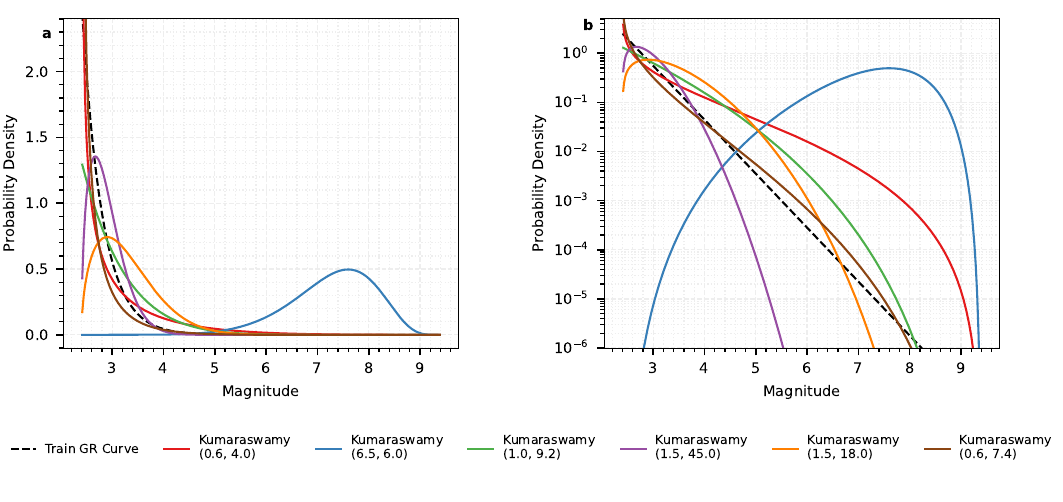}
    \caption{
        \textbf{Examples of Kumaraswamy distributions for different parameter pairs.} Several realizations of the Kumaraswamy distribution (Eq. \ref{eq:kumaraswamy}) are shown on a linear scale (panel \textbf{a}) and on a logarithmic scale (panel \textbf{b}) for the parameter pairs $(a,b)$ indicated in the legend. All examples are plotted on the same support used for the Southern California predictions, i.e., magnitudes $\in[2.4,\,9.4]$. For comparison, the GR distribution fitted to the Southern California training set is also shown.
    }
    \label{fig:kumaraswamy_example}
\end{figure}
\FloatBarrier

\begin{figure}[ht!]
    \centering
    \includegraphics[width=1\textwidth]{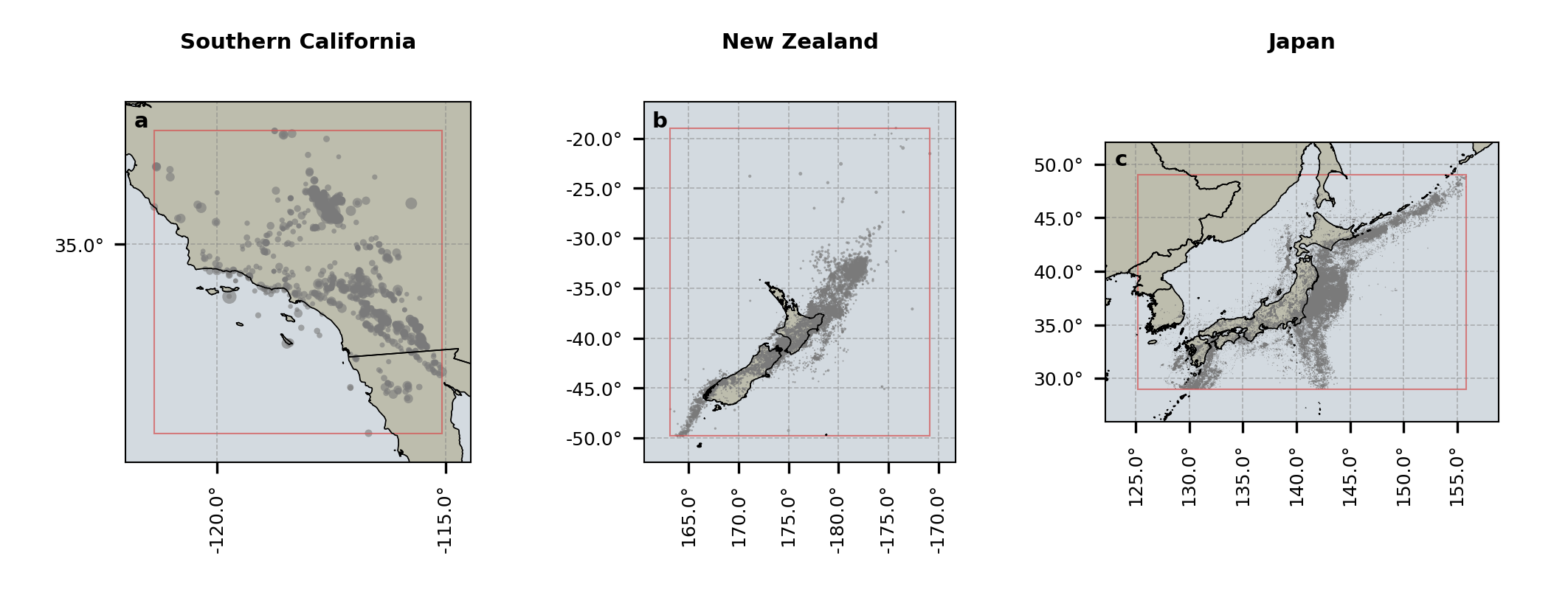}
    \caption{
        \textbf{Tested regions}. Maps of the regions examined in this manuscript. The Southern California, New Zealand and Japan data sets are presented in \textbf{a, b, c} respectively. Grey scatter indicates events in the test set of each region. A red rectangle indicates the selected area for this study.
    }
    \label{fig:tested_regions}
\end{figure}
\FloatBarrier

\begin{figure}[ht!]
    \centering
    \includegraphics[width=\linewidth,keepaspectratio]{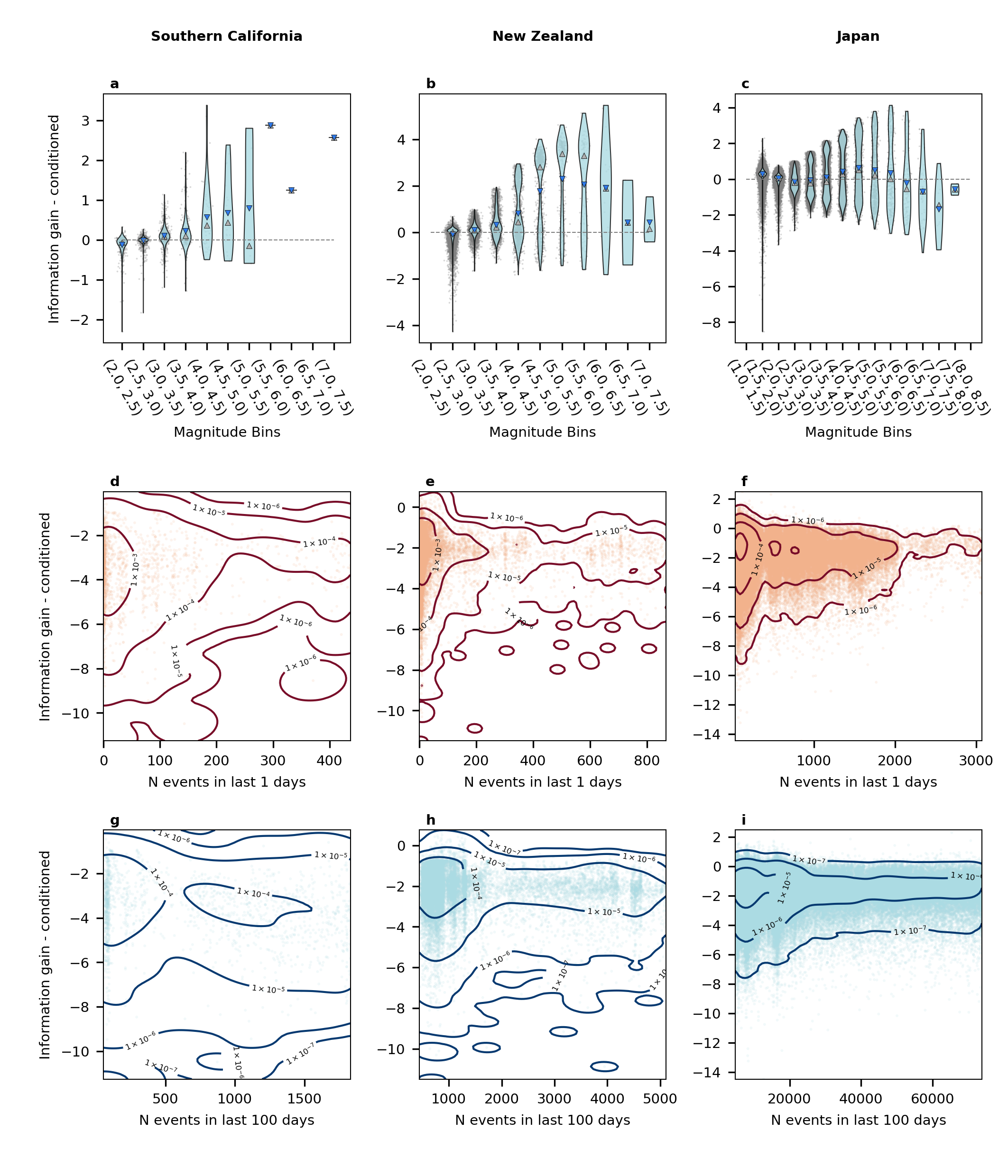}
    \caption{
        \textbf{MAGNET's information gain dependence on magnitude and seismicity}.
        \textbf{a-c)} Distribution of conditioned information gain, $\Delta \ell_i$ (equation \ref{eq:information_gain}), per magnitude bin.
        The distribution of the likelihood difference, i.e., information gain per event over the common GR benchmark, is presented for each 0.5 magnitude interval.
        A positive $\Delta \ell_i$ value indicates an advantage for MAGNET over the benchmark.
        The information gain values are presented by the gray scatter dots, with the mean (median) value per bin indicated by the blue (grey) downward (upward) pointing triangle.
        The shape of the distribution is presented by the blue violin plot.
        \textbf{d-i)} MAGNET’s information gain dependence on seismicity.
        Seismicity (i.e., number of events) in the past day (\textbf{d-f}, red) and past 100 days (\textbf{g-i}, blue) as a function of the information gain of MAGNET ($\Delta \ell_i$) over the common GR benchmark.
        Test set examples are conditioned on temporal incompleteness.
        Light scatter dots indicate values for individual events in the test set, and contour lines indicate the density of the scatter.
        For reference, the Pearson correlation coefficients of the presented data are $0.10,\ -0.05,\ -0.04,\ -0.23,\ -0.06,\ -0.04$ for plots d-i, respectively.
    }
    \label{fig:info_gain_per_bin}
\end{figure}
\FloatBarrier

\begin{figure}[!tbp]
    \centering
    \includegraphics[width=\linewidth,keepaspectratio]{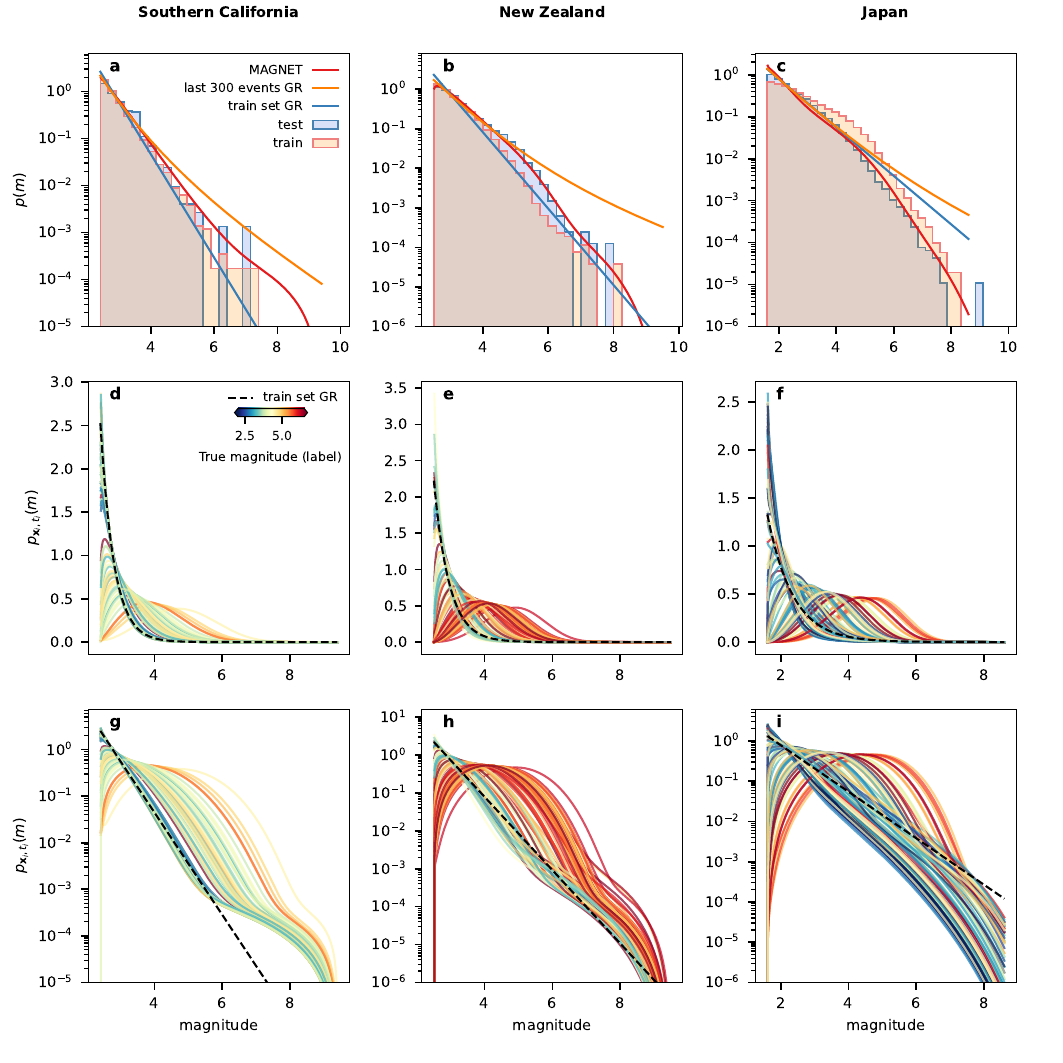}
    \caption{
        \textbf{MAGNET’s output for all data set.}
        \textbf{a-c},
        Marginal magnitude distribution, i.e., stationary p(m), for the Southern California, New Zealand, and Japan data sets, respectively. Histograms of the train and test sets are shown in orange and blue, respectively.
        \textbf{d-f},
        PDFs generated by MAGNET for 100 randomly sampled events from the test sets of the Southern California, New Zealand, and Japan data sets, respectively.
        These events were sampled based on their label magnitudes from an exponential distribution.
        The Gutenberg-Richter distribution fitted on the train set is superimposed (dashed black line, with parameters $(\beta, m_c)$ equal to $(2.53, 2.4), (2.22, 2.5), (1.33, 1.6)$ for Southern California, New Zealand and Japan, respectively).
        The true magnitude label, i.e., the actual magnitude of the event in question, is indicated by the color, as interpreted by the colorbar in d.
        \textbf{g-i}, Same data presented as in d-f, but on a logarithmic scale.
    }
    \label{fig:model_output_em}
\end{figure}
\FloatBarrier

\begin{table}[!tbp]
    \centering
    \includegraphics[width=\linewidth,height=0.6\textheight,keepaspectratio]{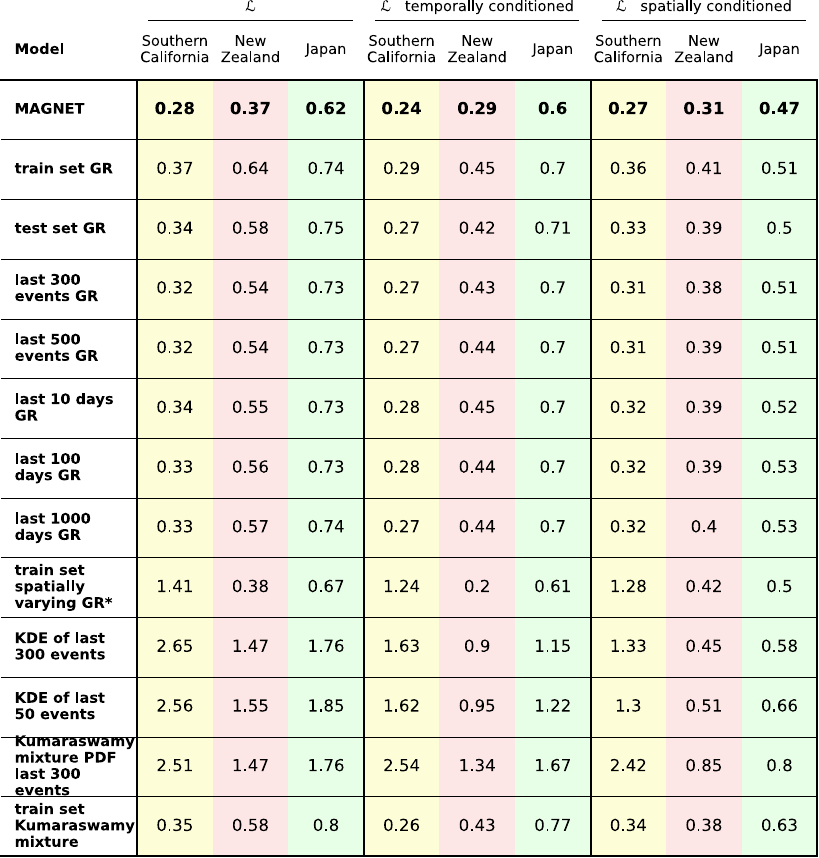}
    \vspace{\baselineskip}
    \caption{Mean score, $\mathcal{L}$, for all tested benchmarks. $\mathcal{L}$ is computed by Eq. \ref{eq:likelihood}. Lower score indicates a better magnitude predictor, best score in column is indicated in bold. First 3 columns display scores for the raw calculation of $\mathcal{L}$, middle and right column triplets display the scores for the temporally and spatially conditioned $\mathcal{L}$ scores, respectively.
        (*) The \textit{train set spatially varying GR} benchmark is evaluated on a smaller subset of events due to methodological  constraints (see Methods); In some cases this might show a superior log likelihood score, but this advantage disappears when evaluated on an identical event set, as shown in Supplementary Table \ref{tab:mean_ll_diff_MAGNET_spatial_GR}.
    }
    \label{tab:mean_ll_all_benchmarks}
\end{table}
\FloatBarrier

\begin{sidewaystable}[!p]
    \centering
    \includegraphics[width=0.88\linewidth]{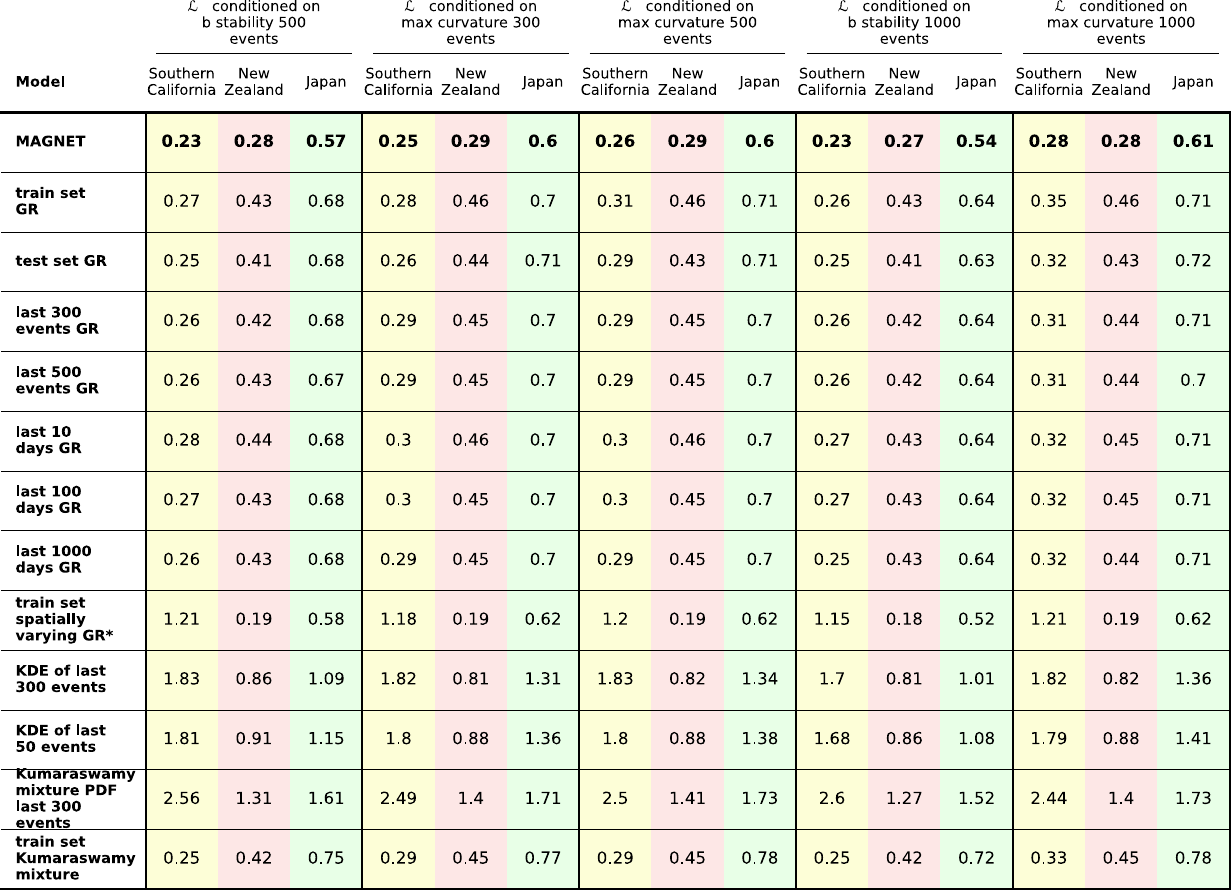}
    \vspace{\baselineskip}
    \caption{Mean score, $\mathcal{L}$, for all tested benchmarks, temporally conditioned by various calculation methods of $m_c(t)$.
        (*) The \textit{train set spatially varying GR} benchmark is evaluated on a smaller subset of events due to methodological  constraints (see Methods); In some cases this might show a superior log likelihood score, but this advantage disappears when evaluated on an identical event set, as shown in Supplementary Table \ref{tab:mean_ll_diff_MAGNET_spatial_GR}.
    }
    \label{tab:mean_ll_additional_mc_cond}
\end{sidewaystable}
\FloatBarrier

\begin{sidewaystable}[!p]
    \centering
    \includegraphics[width=0.88\linewidth]{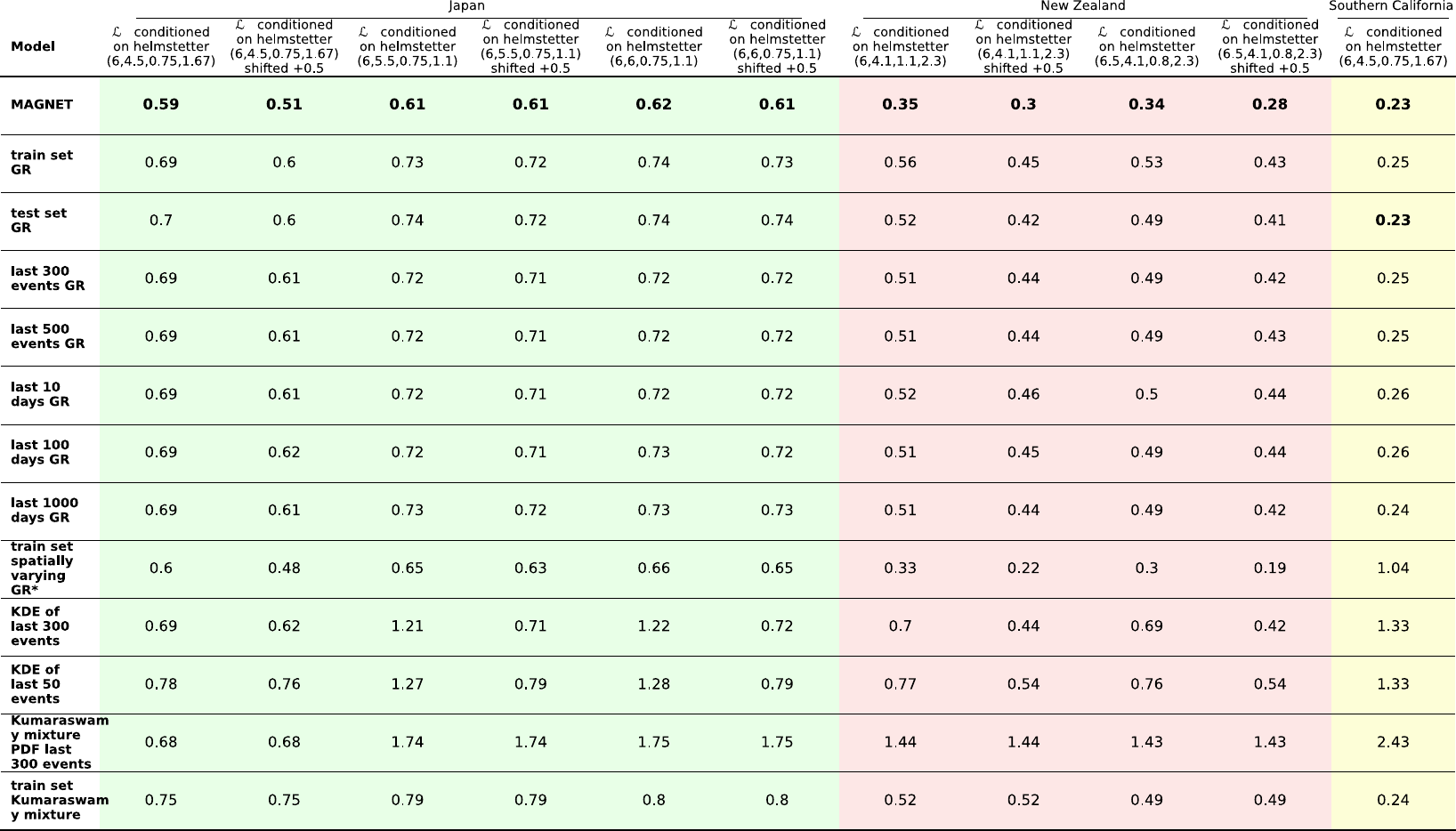}
    \vspace{\baselineskip}
    \caption{Mean score, $\mathcal{L}$, for all tested benchmarks, temporally conditioned on $m_c(t)$ according to the Helmstetter formula for short term incompleteness after a large earthquake: $m_c(t)=\max\{m_i-G-H\log_{10}(t-t_i), m_{min}\}$. where $m_i$ is the magnitude of the main shock, to be considered above a specific threshold
        $m_t$, $G$ and $H$ are fitting parameters and $t-t_i$ is the time since the mainshock, in days.
        The secondary header of the table above indicates the fitting parameters tested according to the key $(m_t, G, H, m_{min})$.
        Where \textit{shifted +0.5} is indicated, a 0.5 magnitude unit was added to the resulting $m_c(t)$.
        (*) The \textit{train set spatially varying GR} benchmark is evaluated on a smaller subset of events due to methodological  constraints (see Methods); In some cases this might show a superior log likelihood score, but this advantage disappears when evaluated on an identical event set, as shown in Supplementary Table \ref{tab:mean_ll_diff_MAGNET_spatial_GR}.
    }
    \label{tab:mean_ll_helm_conditioning}
\end{sidewaystable}
\FloatBarrier

\begin{sidewaystable}[!p]
    \centering
    \includegraphics[width=0.92\linewidth]{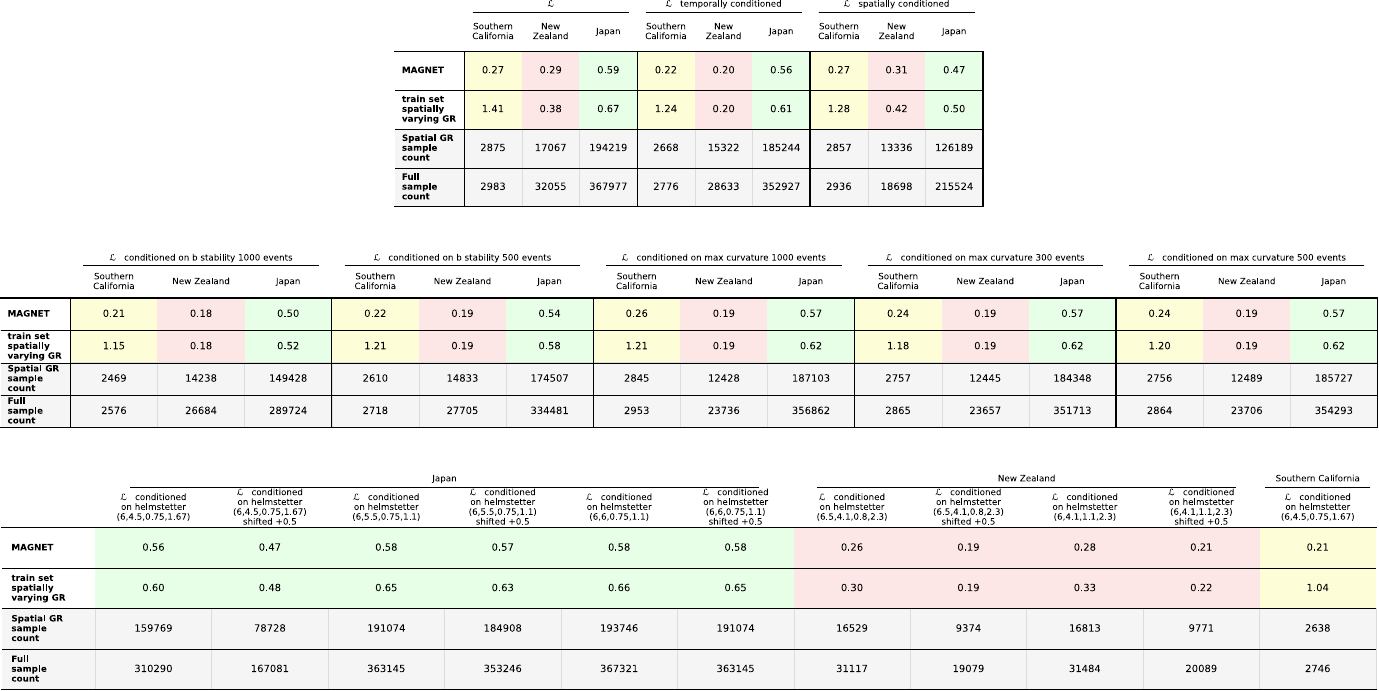}
    \vspace{\baselineskip}
    \caption{Mean score, $\mathcal{L}$, for MAGNET and the benchmark \textit{train set spatially varying GR} computed on a subset of valid events according to Taroni et al. (2021). As statistical reliability is intrinsically linked to sample size, providing these counts is essential for assessing the robustness of the results, particularly given the significant data filtering required for spatial conditioning. The number of events used in the calculation is indicated under 'Spatial GR sample count'. Total number of events for each temporal conditioning $m_c(t)$ method is indicated under 'Full sample count'. Tables correspond to those presented in Supplementary Tables \ref{tab:mean_ll_all_benchmarks}, \ref{tab:mean_ll_additional_mc_cond} and \ref{tab:mean_ll_helm_conditioning}.}

    \label{tab:mean_ll_diff_MAGNET_spatial_GR}
\end{sidewaystable}
\FloatBarrier

\begin{figure}[ht!]
    \centering
    \includegraphics[width=1\textwidth]{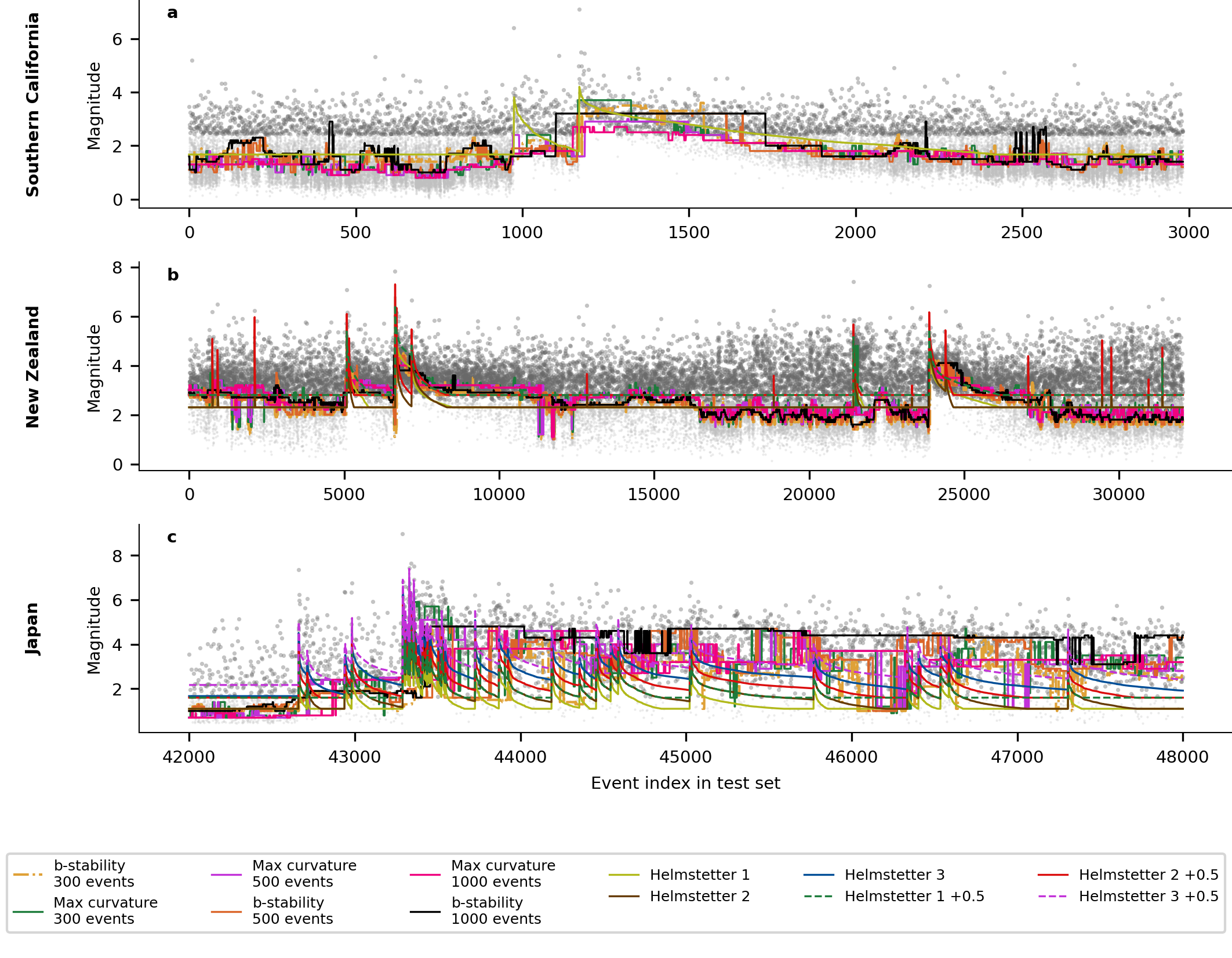}
    \caption{
    \textbf{Temporal Incompleteness for test sets of all regions examined.} Calculated temporal incompleteness by various methods for test sets is presented by the colorful curves, superimposed over the events above (dark grey, included in model evaluation) and below (small, light grey) the completeness magnitude of the train set, $m_c^{(train)}$. Events are plotted by the index of their appearance in the test set. Data for Japan (c) is displayed only around the March 11 2011 T\=ohoku mainshock to reduce data density and enhance clarity of the figure.
    }
    \label{fig:temp_incompleteness}
\end{figure}
\FloatBarrier

\begin{figure}[ht!]
    \centering
    \includegraphics[width=1\textwidth]{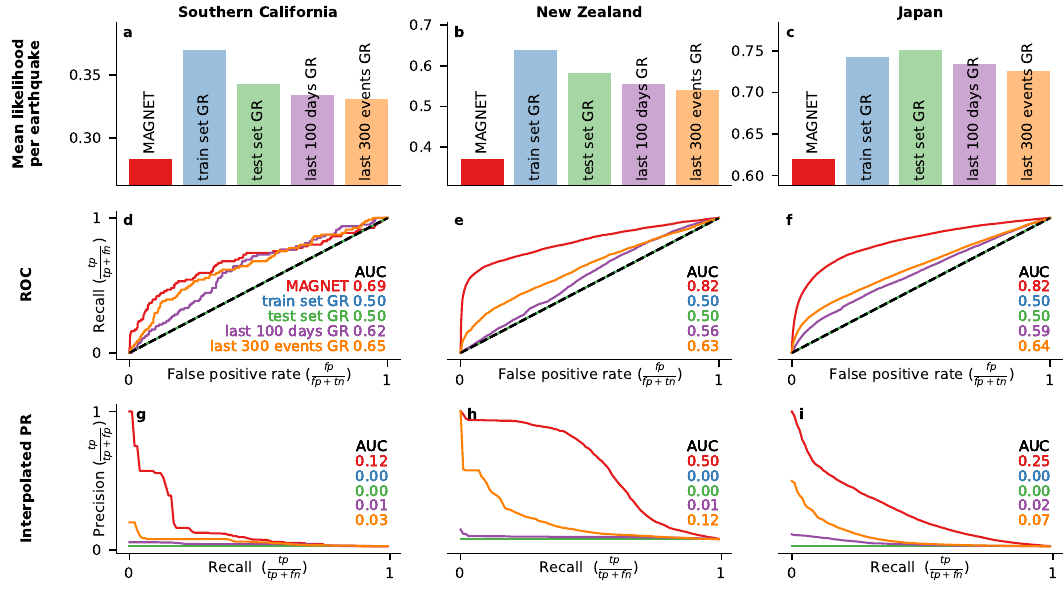}
    \vspace{\baselineskip}
    \caption{
        \textbf{Benchmarking MAGNET.} Comparison of MAGNET's raw performance (non-conditioned) to common benchmarks.
        \textbf{a}, \textbf{b}, \textbf{c}, Minus mean information content the MAGNET model (red) and other common benchmark magnitude predictors (see labels on bars in the figure).
        A lower score indicates a better preforming model.
        \textbf{d, e, f}, Receiver Operating Characteristic (ROC) and \textbf{g, h, i} the interpolated precision-recall (PR) curve for a binary classifier determining the next event will be large ($m>=4$). The performance of such a classifier can be quantified by the area under the curve (AUC), noted in each frame, color coded identically to the bar plots.
        For the AUC metrics, a higher score indicates a better preforming classifier.
    }
    \label{fig:metrics_not_conditioned}
\end{figure}
\FloatBarrier

\begin{figure}[!tbp]
    \centering
    \includegraphics[width=\linewidth,keepaspectratio]{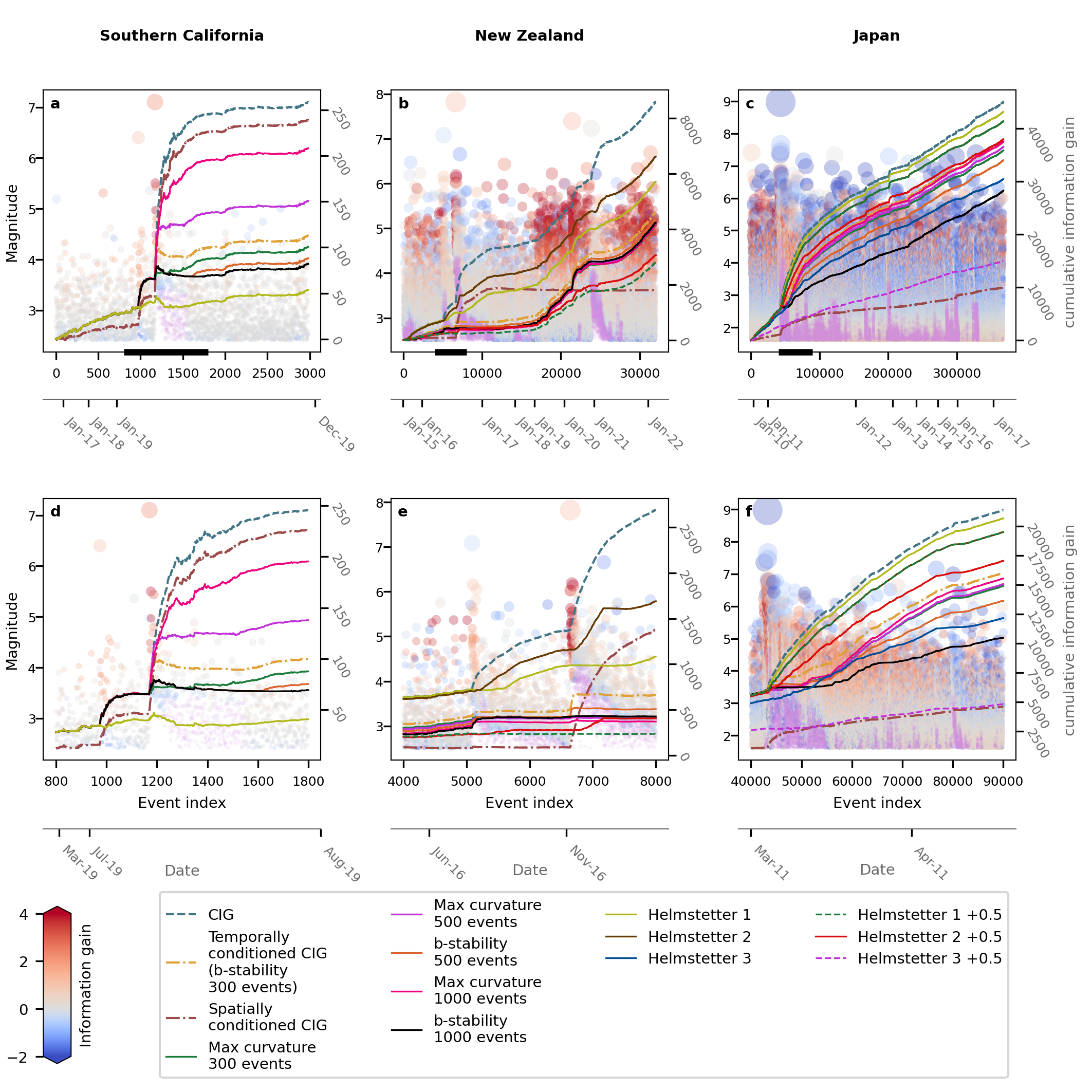}
    \caption{
        \textbf{Information gain of MAGNET, with conditioning on various $m_c(t)$ calculations.} Scattered dots indicate magnitude at event index, colored by the temporally conditioned information gain over the common GR benchmark per event. Events below the temporal incompleteness, i.e. conditioned IG cannot be calculated, are plotted in purple. \textbf{a, b, c} Information gain of individual events in the Southern California, New Zealand and Japan test sets, respectively. Secondary horizontal axis (grey) indicates the corresponding origin time. Cumulative Information gain (CIG) (dashed blue), spatially conditioned CIG (dashed red), and temporally conditioned CIG using various calculation of $m_c(t)$ (see legend) curves are superimposed on the scatter. Details about the various methods of calculation of $m_c(t)$ can be found in the Methods section.
        \textbf{d, e, f} show the same data zoomed in on areas of interest: the 2019 Ridgecrest earthquake sequence in Southern California, the 2016 Te Araroa and 2016 Kaik\=oura earthquakes in New Zealand, and the 2011 T\=ohoku earthquake and some of its aftershocks in Japan.
    }
    \label{fig:info_gain_EM}
\end{figure}
\FloatBarrier

\begin{figure}[ht!]
    \centering
    \includegraphics[width=1\textwidth]{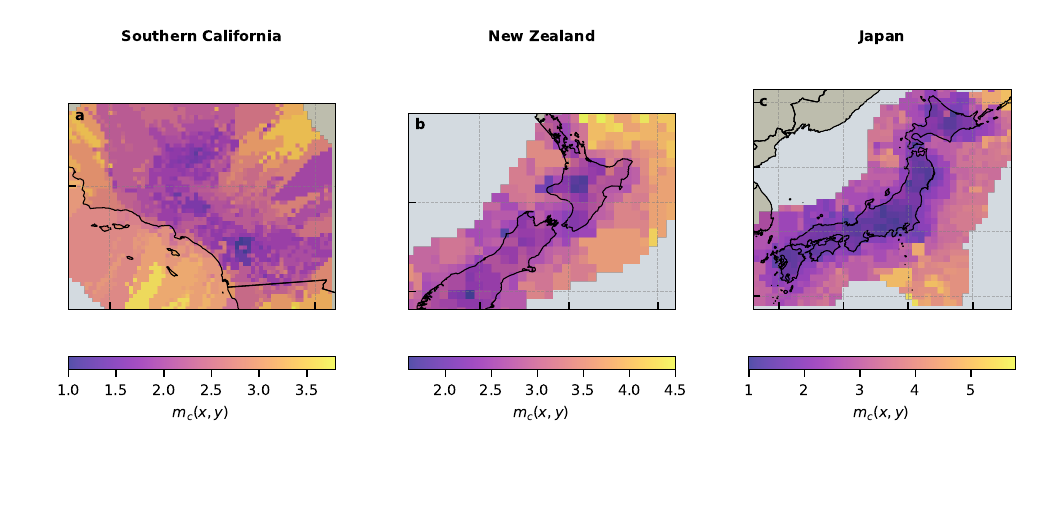}
    \caption{
        \textbf{Spatial Incompleteness for test sets of all regions examined.} The calculation method follows Taroni et al. (2021). Details are given in the Methods section.
    }
    \label{fig:mc_maps}
\end{figure}
\FloatBarrier

\begin{figure}[ht!]
    \centering
    \includegraphics[width=1\textwidth]{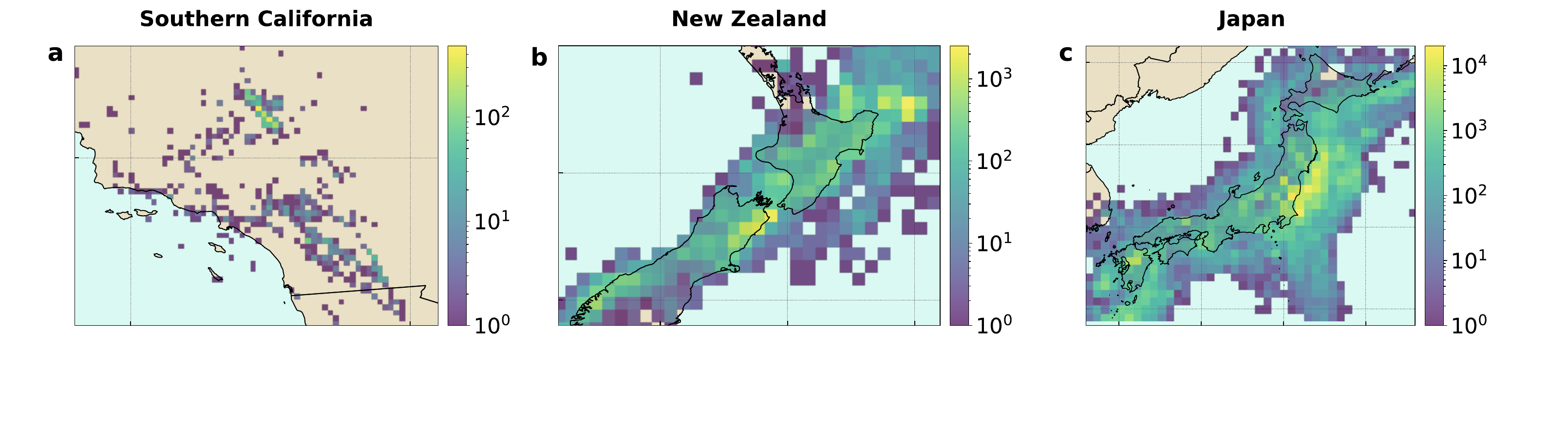}
    \caption{
        \textbf{Event count in examined regions}. Number of events per bin is indicated in the adjacent color bar per inset. Bin divisions are identical to those presented in Figures \ref{fig:info_gain_over_time}e-g.
    }
    \label{fig:seismicity_grid}
\end{figure}
\FloatBarrier

\begin{table}[htbp]
    \centering
    \begin{tabular}{c|c}
        \multicolumn{2}{c}{Recent earthquakes features} \\
              & $f(x, y, t, x_j, y_j, d_j, t_j, M_j)$   \\
        \hline
        $f_1$ & $t - t_j$                               \\
        $f_2$ & $\frac{1}{t + \varepsilon - t_j}$       \\
        $f_3$ & $\log{t + \varepsilon - t_j}$           \\
        $f_4$ & $\exp \left(M_j \right)$                \\
        $f_5$ & $\exp \left(-M_j \right)$               \\
        $f_6$ & $x_j$                                   \\
        $f_7$ & $y_j$                                   \\
        $f_8$ & $d_j$                                   \\
        $f_9$ & $\log{d_j}$                             \\
    \end{tabular}
    \caption{
        \textbf{Features of the \textit{Recent Earthquakes Encoder}}.
        The table presents the list of geophysically inspired functions that are used as an input to the LSTM network. $x, y, t$ are the coordinates and time of the query event, and $x_j, y_j, d_j, t_j, M_j$ are the coordinates, depth, time and magnitude of the $j$~-~th event in the past.
        $\varepsilon$ is a small constant to avoid division by zero.
    }
    \label{tab:recent_earthquake_features}
\end{table}
\FloatBarrier

\end{document}